\documentclass[11pt,onecolumn]{IEEEtran}
\usepackage{amsmath}
\usepackage{amssymb}
\usepackage{amsfonts}
\usepackage{epsfig}
\usepackage{subfigure}
\usepackage{calc}
\usepackage{color}
\usepackage[all]{xy}
\usepackage{bm}
\usepackage{enumerate}
\usepackage{hyperref}
\usepackage[noadjust]{cite}
\usepackage{etoolbox} 
\usepackage{caption}
\makeatletter
\patchcmd{\@makecaption}
{\scshape}
{}
{}
{}
\makeatletter
\patchcmd{\@makecaption}
{\\}
{.\ }
{}
{}
\makeatother

\newtheorem{defn}{Definition}
\newtheorem{thm}{Theorem}[section]
\newtheorem{cor}[thm]{Corollary}
\newtheorem{prop}{Proposition}

\newtheorem{lem}[thm]{Lemma}
\newtheorem{conj}[thm]{Conjecture}
\newtheorem{constr}[thm]{Construction}
\newtheorem{note}{Remark}

\newcommand{\bit}{\begin{itemize}}
\newcommand{\eit}{\end{itemize}}
\newcommand{\bcor}{\begin{cor}}
\newcommand{\ecor}{\end{cor}}
\newcommand{\beq}{\begin{equation}}
\newcommand{\eeq}{\end{equation}}
\newcommand{\beqn}{\begin{equation*}}
\newcommand{\eeqn}{\end{equation*}}
\newcommand{\bea}{\begin{eqnarray}}
\newcommand{\eea}{\end{eqnarray}}
\newcommand{\bean}{\begin{eqnarray*}}
\newcommand{\eean}{\end{eqnarray*}}
\newcommand{\ben}{\begin{enumerate}}
\newcommand{\een}{\end{enumerate}}
\newcommand{\bdefn}{\begin{defn}}
\newcommand{\edefn}{\end{defn}}
\newcommand{\bnote}{\begin{note}}
\newcommand{\enote}{\end{note}}
\newcommand{\bprop}{\begin{prop}}
\newcommand{\eprop}{\end{prop}}
\newcommand{\blem}{\begin{lem}}
\newcommand{\elem}{\end{lem}}
\newcommand{\bthm}{\begin{thm}}
\newcommand{\ethm}{\end{thm}}
\newcommand{\bconj}{\begin{conj}}
\newcommand{\econj}{\end{conj}}
\newcommand{\bconstr}{\begin{constr}}
\newcommand{\econstr}{\end{constr}}
\newcommand{\bpf}{\begin{proof}}
\newcommand{\epf}{\end{proof}}

\begin{document}
\sloppy
\title{Codes for Distributed Storage} 
\author{
 \IEEEauthorblockN{Vinayak Ramkumar\IEEEauthorrefmark{1}, Myna Vajha\IEEEauthorrefmark{1}, S. B. Balaji\IEEEauthorrefmark{2}, M. Nikhil Krishnan\IEEEauthorrefmark{3}, \\ Birenjith Sasidharan\IEEEauthorrefmark{4}, and P. Vijay Kumar\IEEEauthorrefmark{1} \\ \ \\} 
 \IEEEauthorblockA{\IEEEauthorrefmark{1}%
 	Department of Electrical Communication Engineering, IISc Bangalore\\}
   \IEEEauthorblockA{\IEEEauthorrefmark{2}%
 	Qualcomm, Bangalore\\}
 \IEEEauthorblockA{\IEEEauthorrefmark{3}%
 	Department of Electrical and Computer Engineering, University of Toronto\\}
 \IEEEauthorblockA{\IEEEauthorrefmark{4}%
Department  of  Electronics  and  Communication  Engineering,    GEC Barton Hill, Trivandrum\\}
\thanks{This survey article will appear as a chapter in the upcoming "A Concise Encyclopedia of Coding Theory", W. C. Huffman, J.-L. Kim, and P. Sol\'e, CRC Press. 
	
P. V. Kumar is also a Visiting Professor at the University of Southern California.  His  research  is  supported  in  part  by  the J C Bose National Fellowship JCB/2017/000017 and  in  part  by the  NetApp University Research Fund SVCF-0002.
}
}
\maketitle

\begin{abstract}
This chapter deals with the topic of designing reliable and efficient codes for the storage and retrieval of large quantities of data over storage devices that are prone to failure.    For long, the traditional objective has been one of ensuring reliability against data loss while minimizing storage overhead.   More recently, a third concern has surfaced, namely of the need to efficiently recover from the failure of a single storage unit, corresponding to recovery from the erasure of a single code symbol.  We explain here, how coding theory has evolved to tackle this fresh challenge. 
\end{abstract}

\section{Introduction}

The traditional means of ensuring reliability in data storage is to store multiple copies of the same file in different storage units. Such a replication strategy is clearly inefficient in terms of storage overhead. By \textbf{storage overhead}, we mean the ratio of the amount of data stored pertaining to a file to the size of the file itself. Modern-day data centers can store several Exabytes of data and there are enormous costs associated with such vast amounts of storage not only in terms of hardware, software and maintenance, but also in terms of the power and water consumed.  For this reason, reduction in storage overhead is of paramount importance.   An efficient approach to reduce storage overhead, without compromising on reliability,  is to employ erasure codes instead of replication. When an $[n,k]$ erasure code is employed to store a file, the file is first broken into $k$ fragments.  To this, $n-k$ redundant fragments are then added and the resultant $n$ fragments are stored across $n$ storage units. Thus the storage overhead incurred is given by $\frac{n}{k}$. In terms of minimizing storage overhead, a class of codes known as Maximum Distance Separable (MDS) codes, of which Reed-Solomon (RS) codes are the principal example, are the most efficient. MDS codes have the property that the entire file can be obtained by connecting to any $k$ storage units, which means that an MDS codes can handle the failure of $n-k$ storage units without suffering data loss. RS codes are widely used  in current-day distributed-storage systems. Examples include, the $[9,6]$ RS code in Hadoop Distributed File System with Erasure Coding (HDFS-EC), the $[14,10]$ RS code employed in Facebook's f4 BLOB storage and the $[12,8]$ RS code employed in Baidu's Atlas cloud storage \cite{DauDuuKiaMil}. 

A second important consideration governing choice of erasure code is the ability of the code to efficiently handle the failure of an individual storage unit, as such failures are a common occurrence  \cite{SathiaAstPap_Xorbas,RashmiShahGuKuang}. 
From here on, we will interchangeably refer to a storage unit as a node, as we will at times, employ a graphical representation of the code. We use the term node failure to encompass not only the actual physical failure of a storage unit, but also instances where a storage unit is down for maintenance or else is unavailable on account of competing serving requests. 
Thus, the challenge is to design codes which are efficient not only with respect to storage overhead, but which also having the ability to efficiently handle the repair of a failed node. The nodes from which data is downloaded for the repair of a failed node are termed as helper nodes and the number of helper nodes contacted is termed the \textbf{repair degree}. The amount of data downloaded to a replacement node from the helper nodes during node repair is called the \textbf{repair bandwidth}.  

If the employed erasure code is an $[n,k]$ MDS code, then the conventional approach towards repair of a failed node is as follows.  A set of $k$ helper nodes would be contacted and the entire data from these $k$ nodes would then be downloaded.  This would then permit reconstruction of the entire data file and hence in particular, enable repair of the failed node.  Thus under this approach, the repair bandwidth is $k$ times the amount of data stored in the replacement node, which is clearly wasteful of network bandwidth resources and can end up clogging the storage network.

\begin{center}
	\begin{figure}[h!]
		\centering
		\includegraphics[width=5in]{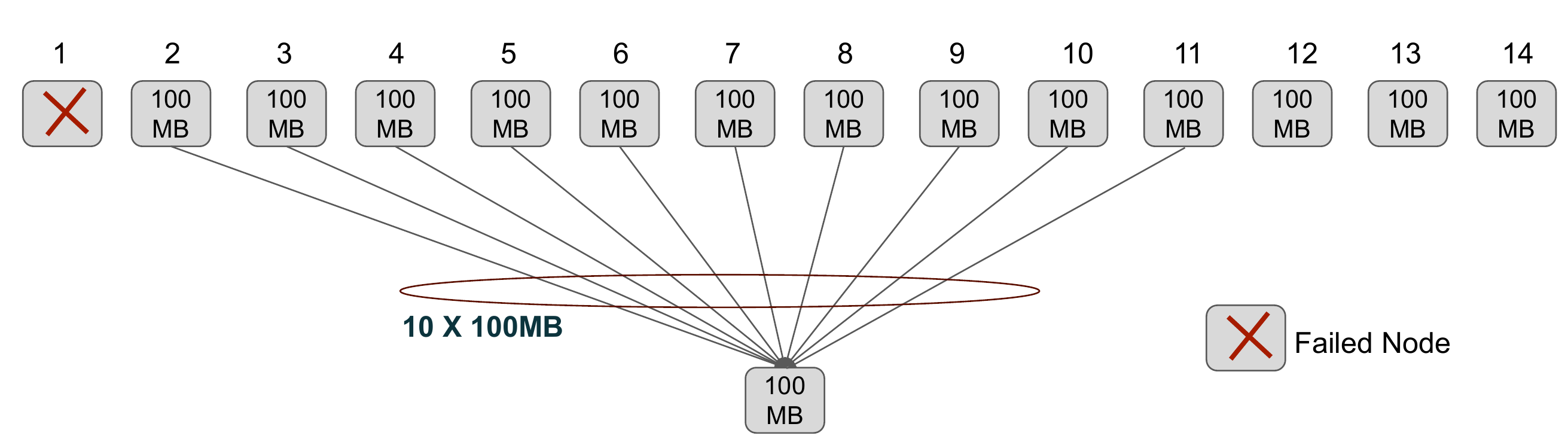}
			\caption{Conventional repair of a failed node in $[14,10]$ RS code.}   
		\label{fig:rs_14_10}    
	\end{figure}      
\end{center} 

This is illustrated in Fig.~\ref{fig:rs_14_10} in the case of Facebook's $[14,10]$ RS code.  If each node stores $100$MB of data, then the total amount of data download needed for the repair of a node storing $100$MB from the $10$ helper nodes equals  $1$GB. Thus in general, in the case of an $[n,k]$ MDS code, the repair degree equals $k$ and repair bandwidth is the entire file size which is $k$ times the amount of data stored in a node. As will be seen, under conventional node repair, MDS codes are inefficient both in terms of repair degree and repair bandwidth.  

Coding theorists responded to the challenge of node repair by coming up with two classes of codes, known as {\em ReGenerating Codes} (RGCs) \cite{DimGodWuWaiRam} and {\em Locally Recoverable Codes} (LRCs) \cite{GopHuaSimYek}. The goal of an RGC is to reduce the repair bandwidth, whereas with an LRC, one aims to minimize the repair degree. 
Recently much progress has also been made in coming up with more efficient approaches for the repair of RS codes \cite{ShanPapDimCai,GuruWoot}. This chapter will primarily focus on RGC and LRC. We will also briefly discuss novel methods of RS repair, as well as a class of codes known as {\em Locally ReGenerating Codes} (LRGCs), that combine the desirable properties of RGCs and LRCs.

A brief overview of the topic of codes for distributed storage, and from the perspective of the authors, is presented here. For further details, the readers are referred to surveys such as can be found in \cite{BalNikVajSurvey,DimRamWuSuhSurvey,DattaOggSurvey,LiLiSurvey,liuOggSurvey}.  We would like to thank the editors of Advanced Computing and Communications for permitting  the reuse of some material from \cite{ACCS}. 

We begin with a brief primer on RS codes.

\section{Reed-Solomon Code} \label{sec:RS} 

We provide here a brief description of an $[n,k]$ RS code \footnote{Throughout this chapter the term `Reed-Solomon codes' will include what are often called `generalized Reed-Solomon codes', defined specifically in Section \ref{Sec:RS_repair}.}.
Let $a_i \in \mathbb{F}_q$, $i = 0,1,\ldots,k-1$, represent the $k$ message symbols.  Let $x_0,x_1,\ldots,x_{n-1}$  be an arbitrary collection of $n$ distinct elements from $\mathbb{F}_q$ and the polynomial $f$ be defined by: 
\begin{eqnarray*}
	f(x) & = & \sum_{i=0}^{k-1} a_i  \ \prod^{k-1}_{\begin{array}{c} j=0 \\ j \neq i \end{array}}  \frac{(x-x_j)}{(x_i-x_j)}  \ \ := \ \sum_{i=0}^{k-1} b_i x^i .
\end{eqnarray*}
It follows that $f$ is the interpolation polynomial of degree $(k-1)$ that satisfies 
\begin{eqnarray*}
	f(x_i) \ =  \ a_i, & & 0 \leq i \leq (k-1). 
\end{eqnarray*}
\begin{center}
	\begin{figure}[h!]
		\centering
		\includegraphics[width=2.5in]{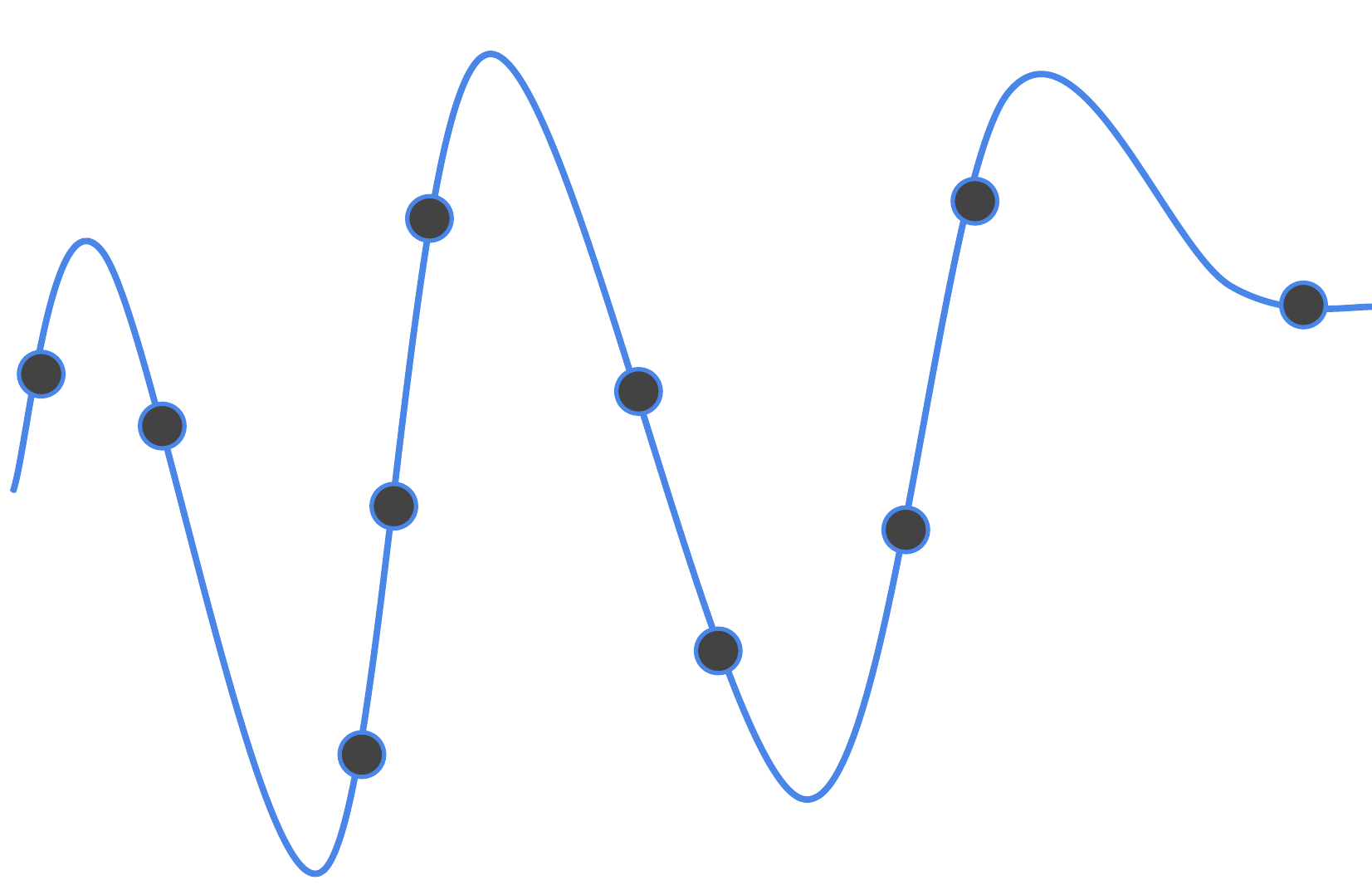} 
				\caption{Illustrating the operating principle of an RS code.}
		\label{fig:RS_code_idea}    
	\end{figure}      
\end{center}  
The $n$ code symbols in the RS codeword corresponding to message symbols $\{a_i\}_{i=0}^{k-1}$ are precisely the $n$ ordered symbols $(f(x_0),f(x_1), \ldots ,f(x_{n-1}))$, see Fig.~\ref{fig:RS_code_idea} for an illustration.  Of these, the first $k$ symbols are message symbols while the remaining are redundant symbols.  

The RS code derives its MDS property from the fact that the polynomial $f$ can be determined from the knowledge of any $k$ evaluations, simply by solving a nonsingular set of $k$ equations in the $k$ unknown coefficients $\{b_i\}_{i=0}^{k-1}$ as shown below 

\begin{eqnarray*}
	\left[ \begin{array}{c}
		f(x_{i_1}) \\ f(x_{i_2}) \\ \vdots \\ f(x_{i_k}) \end{array} \right] & = & 
	\underbrace{\left[ \begin{array}{cccc}
			1 & x_{i_1}& \cdots & x_{i_1}^{k-1}  \\ 
			1 & x_{i_2} & \cdots & x_{i_2}^{k-1}  \\ 
			\vdots & \vdots & \vdots & \vdots \\ 
			1 & x_{i_k} & \cdots & x_{i_k}^{k-1}  
		\end{array} \right] }_{\begin{array}{c} \text{a Vandermonde matrix} \\ \text{and therefore invertible} \end{array} }
	\left[ \begin{array}{c}
		b_0\\  \\ \vdots \\ b_{k-1} \end{array} \right],
\end{eqnarray*}
where $i_1, \ldots, i_k$ are any set of $k$ distinct indices drawn from $\{0, \ldots, n-1\}$. From this it follows that an RS code can recover from the erasure of any $n-k$ code symbols. The conventional approach of repairing a failed node corresponding to code symbol $f(x_j)$ would be to use the contents of any $k$ nodes (i.e., any $k$ code symbols) to recover the polynomial $f$ and then evaluate the polynomial $f$ at $x_j$ to recover the value of $f(x_j)$.

\section{Regenerating Codes}

Traditional erasure codes are scalar codes, i.e., each code symbol corresponds to a single symbol from a finite field.   It turns out however, that the design of codes with improved repair bandwidth calls for codes that have an underlying vector alphabet. Thus, each code symbol is now a vector. The process of moving from a scalar symbol to a vector symbol is referred to as \textbf{sub-packetization}.  The reason for this choice of terminology is that we view a scalar symbol over the finite field $\mathbb{F}_{q^{\alpha}}$ of size $q^{\alpha}$ as being replaced by a vector of $\alpha$ symbols drawn from $\mathbb{F}_q^{\alpha}$. An example is presented in Fig.~\ref{fig:vectorize}. 

\subsection{An Example Regenerating Code and Sub-packetization} 
\begin{figure}[h!]
	\begin{center}
		\includegraphics[width=0.85\textwidth]{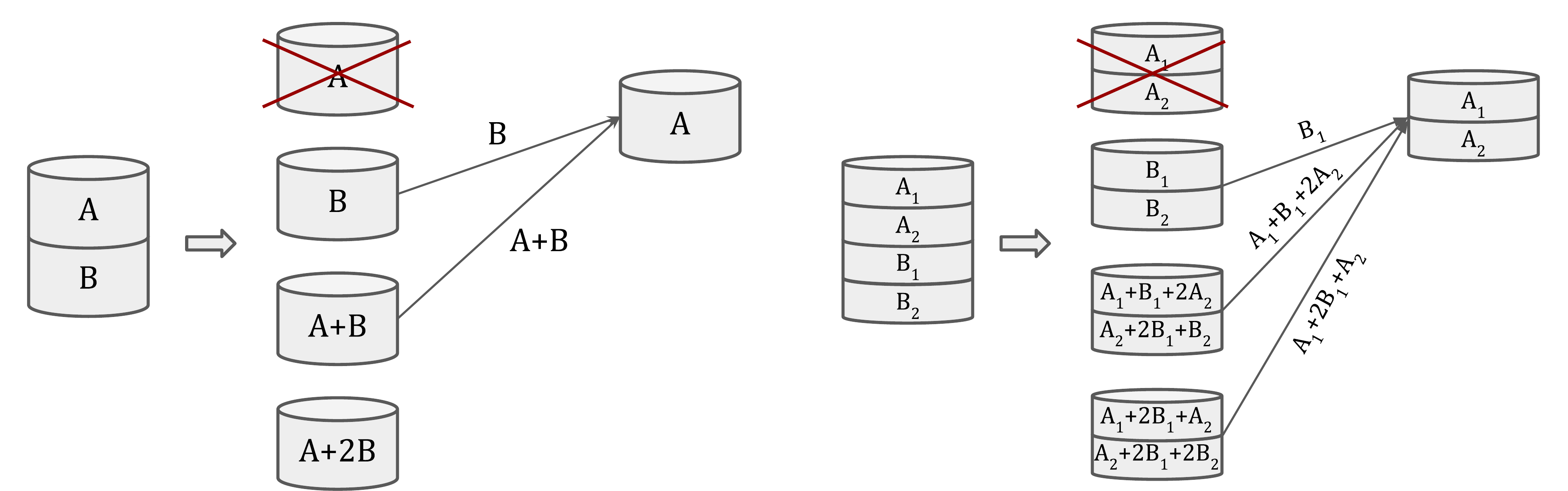}
			\caption{Showing how breaking up a single scalar symbol into two smaller symbols helps improve the repair bandwidth. This breaking up of a symbol is referred to as sub-packetization.  The sub-packetization level equals $2$ here.}
			\label{fig:vectorize} 
	\end{center}
\end{figure}
In Fig.~\ref{fig:vectorize}, the setup on the left represents a $[4,2]$ MDS code.  The symbols stored in the $4$ nodes are respectively, $A,B,A+B,A+2B$, all drawn from a finite field of suitable size, for example $\mathbb{F}_{3^2}$.  To repair the failed node (disk) $1$ that previously stored $A$, we have to download $2$ symbols. Consider next, the setup to the right.  Here the sub-packetization level is $2$, each symbol drawn from $\mathbb{F}_{3^2}$, is replaced by $2$ `half-symbols', each drawn from $\mathbb{F}_3$. Thus $A$ is replaced by $A_1,A_2$, $B$ by $B_1,B_2$.  Note that if the data stored in the remaining two parity nodes is as shown in the figure, then node $1$ can be repaired by downloading $3$ half-symbols in place of two full symbols, thereby achieving a reduction in repair bandwidth. Note however that  we have contacted all the remaining nodes, $3$ in this case, as opposed to $k=2$ in the case of the MDS code. Thus while regenerating codes reduce the repair bandwidth, they do in general, result in increased repair degree. 

\subsection{General Definition of a Regenerating Code} 
\begin{figure}[ht!]
	\vspace*{-0.1in} 
	\begin{center}
		\begin{minipage}{2in}
			\begin{center}
				\includegraphics[trim= 0in  0in 0in 0in, width=2in]{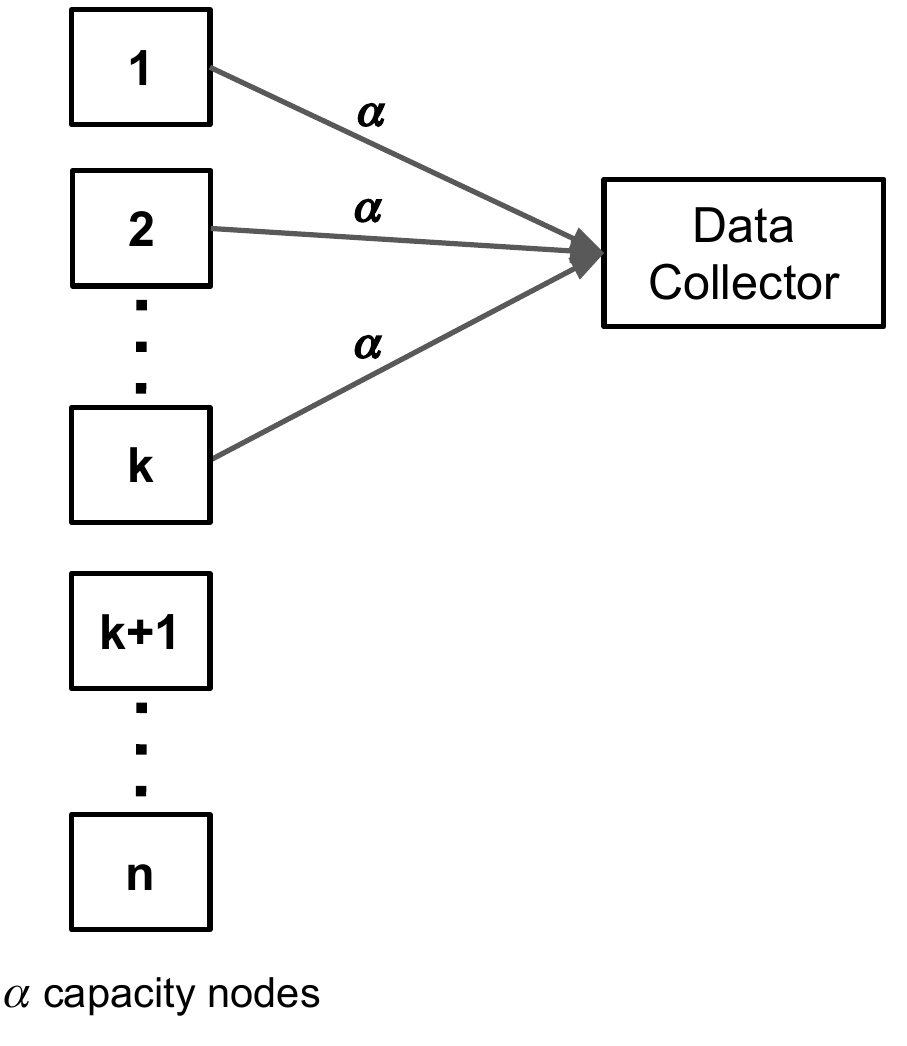} 
			\end{center}
		\end{minipage} \hspace*{0.1in} 
		\begin{minipage}{2in}
			\begin{center}
				\includegraphics[trim= 0in  0in 0in 0in, width=2.2in]{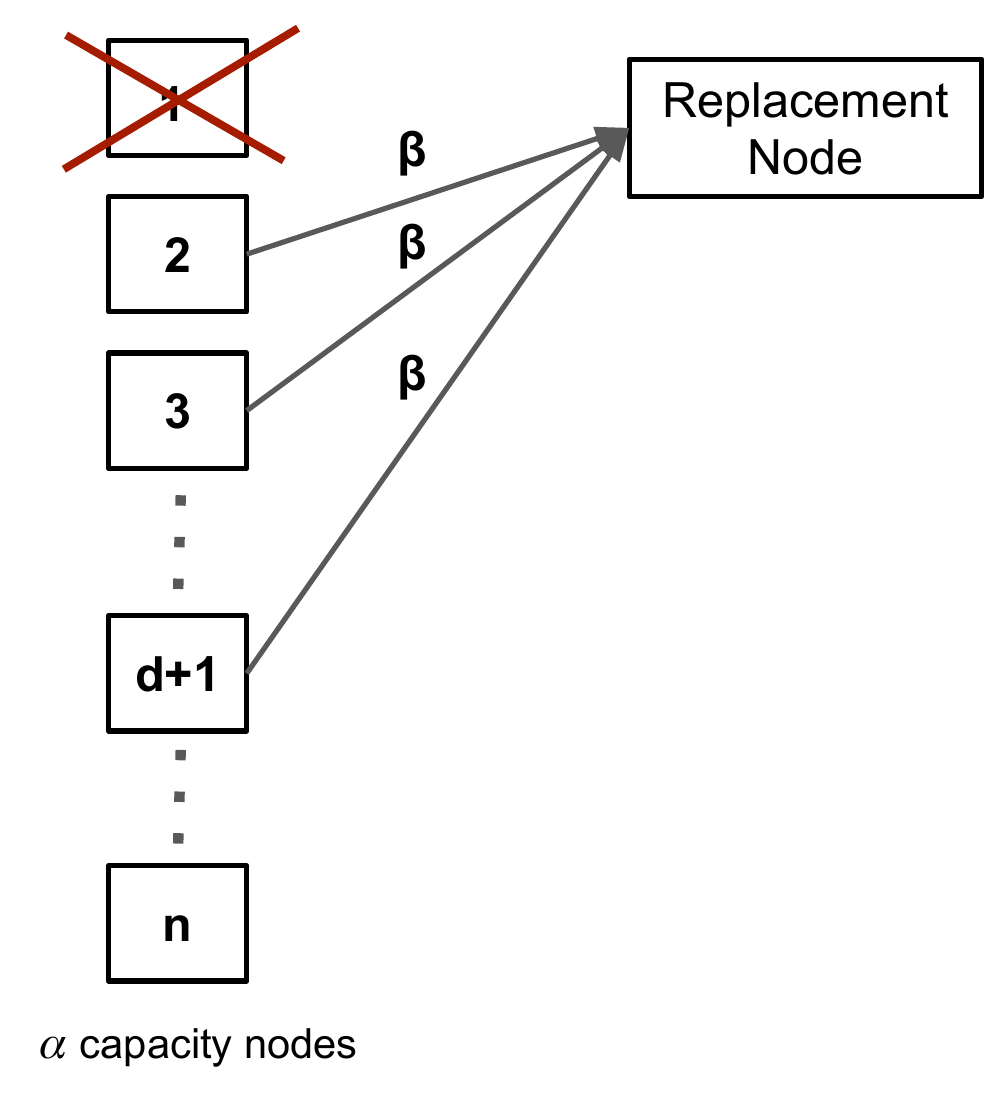} 
			\end{center}
		\end{minipage}
	\end{center}
	\caption{Data collection and node repair properties of an  $\{ (n,k,d), \ (\alpha, \beta), \ B, \ \mathbb{F}_q \}$ regenerating code.}
		\label{fig:RGC}	
\end{figure}

\begin{defn}[\cite{DimGodWuWaiRam}]
	{\em 
		Given a file of size $B$, as measured in number of symbols over $\mathbb{F}_q$, an $\{(n,k,d),(\alpha,\beta), B ,\mathbb{F}_q\}$ {\bf  regenerating code} ${\cal C}$ stores data pertaining to this file across $n$ nodes, where each node stores  $\alpha$ symbols from  the field $\mathbb{F}_q$.
		The code ${\cal C}$ is required to have the following properties (see Fig.~\ref{fig:RGC}): 
		\begin{enumerate}
			\item[i)] 	\textbf{Data Collection:} The entire file can be obtained by downloading contents of any $k$ nodes.
			\item[ii)] \textbf{Node Repair: } If a node fails, then the replacement node can connect to any subset of $d$ helper nodes, where $k \le d \le n-1$, download $\beta$ symbols from each of these helper nodes and repair the failed node.
	\end{enumerate}}
\end{defn}

With respect to the definition above, there could be two interpretations as to what it means to repair a failed node.  One interpretation of repair is that the failed node is replaced by a replacement node which upon repair, stores precisely the same data as did the failed node.  This is called \textbf{Exact Repair} (E-R). This is not however, a requirement and there is an alternative, more general definition of node repair.  Under this more general interpretation, the failed node is replaced by a replacement node in such a way that the resultant collection of $n$ nodes, once again satisfies the requirements of a regenerating code.  Such a replacement of a failed node is termed as \textbf{Functional Repair} (F-R).  
Clearly, E-R is a special case of F-R.  E-R is preferred in practice as it simplifies management of the storage network. 

It is easily verified that the storage overhead of a regenerating code equals $\frac{n\alpha}{B}$, the repair degree is $d$ and the repair bandwidth is $d\beta$. The parameter $\beta$ is typically much smaller than $\alpha$, resulting in savings in repair bandwidth. The parameter $\alpha$ is called the sub-packetization level of the regenerating code and having low sub-packetization level is preferred from a practical perspective (see \cite{VajRamPur_Clay}).
A regenerating code is said to possess the \textbf{optimal-access}  property if no computation is required at the helper nodes during node repair. If repair of a failed node can be done with out any computation at either the helper nodes or the replacement node, then the regenerating code is said to have the \textbf{Repair-By-Transfer} (RBT) property.

\subsection{Bound on File Size} 

It turns out by using the cut-set bound of network coding one can show that the size $B$ of the file encoded by regenerating code must satisfy the following inequality \cite{DimGodWuWaiRam}:
\begin{eqnarray}
\label{eq:cutsetbound}
B \le \sum\limits_{i=0}^{k-1} \min \{\alpha, (d-i)\beta\},
\end{eqnarray}

which we will refer to as the cut-set bound.  This bound holds for F-R (and hence also for E-R). 

\begin{figure}[ht!]
	\begin{center}
		\includegraphics[width=4.5in]{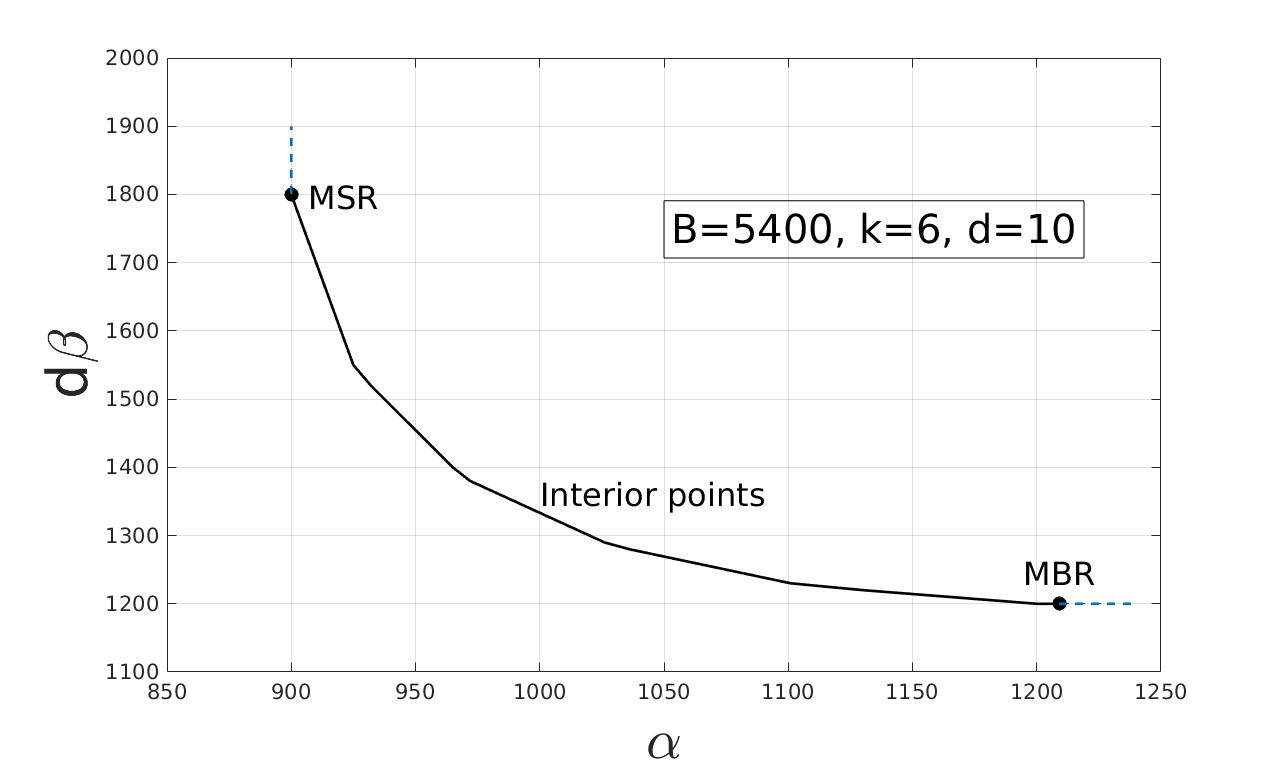}
		\caption{S-RB tradeoff for F-R with $(B=5400,k=6,d=10)$}
				\label{fig:tradeoff}
		\end{center}
\end{figure}

A regenerating code is said to be optimal if the cut-set bound is satisfied with equality and if in addition, decreasing either $\alpha$ or $\beta$ would cause the bound to be violated. There are many flavors of optimality in the sense that for a fixed $(B,k,d)$, inequality \eqref{eq:cutsetbound} can be met with equality by several different pairs $(\alpha, \beta)$. The parameter $\alpha$ determines the storage overhead $\frac{n\alpha}{B}$, whereas $\beta$ is an indicator of normalized repair bandwidth $\frac{d\beta}{B}$. The various pairs of $(\alpha, \beta)$ which satisfy the cut-set bound with equality present a tradeoff between storage overhead and normalized repair bandwidth. The existence of codes achieving the cut-set bound for all possible parameters is known again from network coding in the F-R case.  Thus in the F-R case, the storage-repair bandwidth (S-RB) tradeoff is fully characterized.  
An example of this tradeoff is presented for the case $(B=5400,k=6,d=10)$.  The discussion above suggests that the tradeoff is a discrete collection of optimal pairs $(\alpha, \beta)$.  However, in the plot, these discrete pairs are connected by straight lines, giving rise to the piecewise linear graph in Fig.~\ref{fig:tradeoff}.  The straight line connections have a physical interpretation and correspond to a space sharing solution to the problem of node repair and the reader is referred to \cite{ShaRasKumRam_rbt,TiaSasAggVaiKum} for details. 

\subsection{MSR and MBR Codes\label{subsec:msr_mbr}}

There are two extreme choices of $(\alpha, \beta)$ in this tradeoff. In the first, $\alpha$ is the least possible, as a result of which $\alpha \le (d-k+1)\beta$, thereby forcing $B= k\alpha$.  Under this condition $\beta$ is minimized by setting $\alpha=(d-k+1)\beta$. Regenerating codes with this flavor of optimality are called \textbf{Minimum Storage Regenerating} (MSR) codes.  In the case of an MSR code, the size of the code is given by $q^B=q^{k\alpha}=(q^\alpha)^k=(q^\alpha)^{(n-d_{min}+1)}$, where $d_{min}$ is the minimum distance. It follows that MSR codes achieve the Singleton bound on code size over the vector alphabet $\mathbb{F}_q^\alpha$ and hence MSR codes belong to the class of MDS codes. 
At the other extreme, we have the case $\alpha \ge d\beta$, and file size $B = \sum_{i=0}^{k-1} (d-i)\beta = \  kd\beta - {k \choose 2}\beta$. Here $\alpha$ is minimized by setting $\alpha=d\beta$.  Regenerating codes having these parameters are termed as \textbf{Minimum Bandwidth Regenerating} (MBR) codes. MBR codes have the minimum possible repair bandwidth, but are not MDS. The storage overhead of an MBR code can be shown lower bounded by $2$, whereas MSR codes can have storage overhead arbitrarily close to $1$. 


Several constructions of E-R MSR \cite{RasShaKum_pm11,SuhRam,TamWanBru,SasAgaKum,RawKoyVis_msr,GopFazVar,YeBar_1,YeBar_2,SasVajKum_arxiv,LiTangTian,VajBalKum} and MBR \cite{RasShaKumRam_allerton09,RasShaKum_pm11,LinChungNovelRBTMBR14,KrishKumMBRRep16} codes can be found in the literature.  A selected subset of these constructions are discussed here.

\subsection{Storage-Bandwidth Tradeoff for Exact-Repair}
The cut-set bound \eqref{eq:cutsetbound} may be not achievable under E-R and hence the S-RB tradeoff for E-R may vary from that for F-R.  For brevity, we will refer to the S-RB tradeoff in the case of E-R as the E-R tradeoff and similarly F-R tradeoff in the case of functional repair.  Since E-R codes are a special case of F-R codes, the file size $B$ in the case of E-R case cannot exceed \eqref{eq:cutsetbound}. Constructions of exact-repair MSR and MBR codes are known and hence the E-R tradeoff coincides with the F-R tradeoff at the MSR and MBR points.  Points on the S-RB tradeoff, other than the MSR and MBR points are referred to as interior points. In \cite{ShaRasKumRam_rbt}, it was shown that it is impossible to achieve, apart from one exceptional subset, any interior point of the F-R tradeoff using E-R.  The exceptional set of interior points correspond to a small region of the F-R tradeoff curve, adjacent to the MSR point, see Theorem~\ref{thm:shah_non_exist} below. 

\begin{thm} \label{thm:shah_non_exist} For any given $(n,k\ge 3,d)$, E-R codes do not exist for $(\alpha, \beta, B)$ corresponding to an interior point on the F-R tradeoff, except possibly for a small region in the neighborhood of MSR point with $\alpha$ values in the range, 
	\begin{equation*} \label{eq:exception}
	(d-k+1)\beta  \leq \ \alpha \ \leq (d-k+2)\beta - \left(\frac{d-k+1}{d-k+2}\right) \beta.
	\end{equation*}
	
\end{thm}

This theorem does not eliminate the possibility of E-R codes approaching the F-R tradeoff asymptotically i.e., in the limit as $B \rightarrow \infty$.  The E-R tradeoff for the $(n,k,d)=(4,3,3)$ case was characterized in \cite{Tia} where the impossibility of approaching the F-R tradeoff under E-R was established.  The analogous result in the general case, was established in \cite{SasSenKum_isit}.  Examples of interior-point RGC constructions include layered codes \cite{TiaSasAggVaiKum}, improved layered codes \cite{SenSasKum_ita}, determinant codes \cite{ElyMoh}, cascade codes \cite{ElaMoh_cascade} and multi-linear-algebra-based codes \cite{DurLiWan}.

\subsection{Polygon MBR Code} \label{sec:Polygon_MBR}

\begin{center}
	\begin{figure}[ht!]
		\centering
		\includegraphics[width=2.3in]{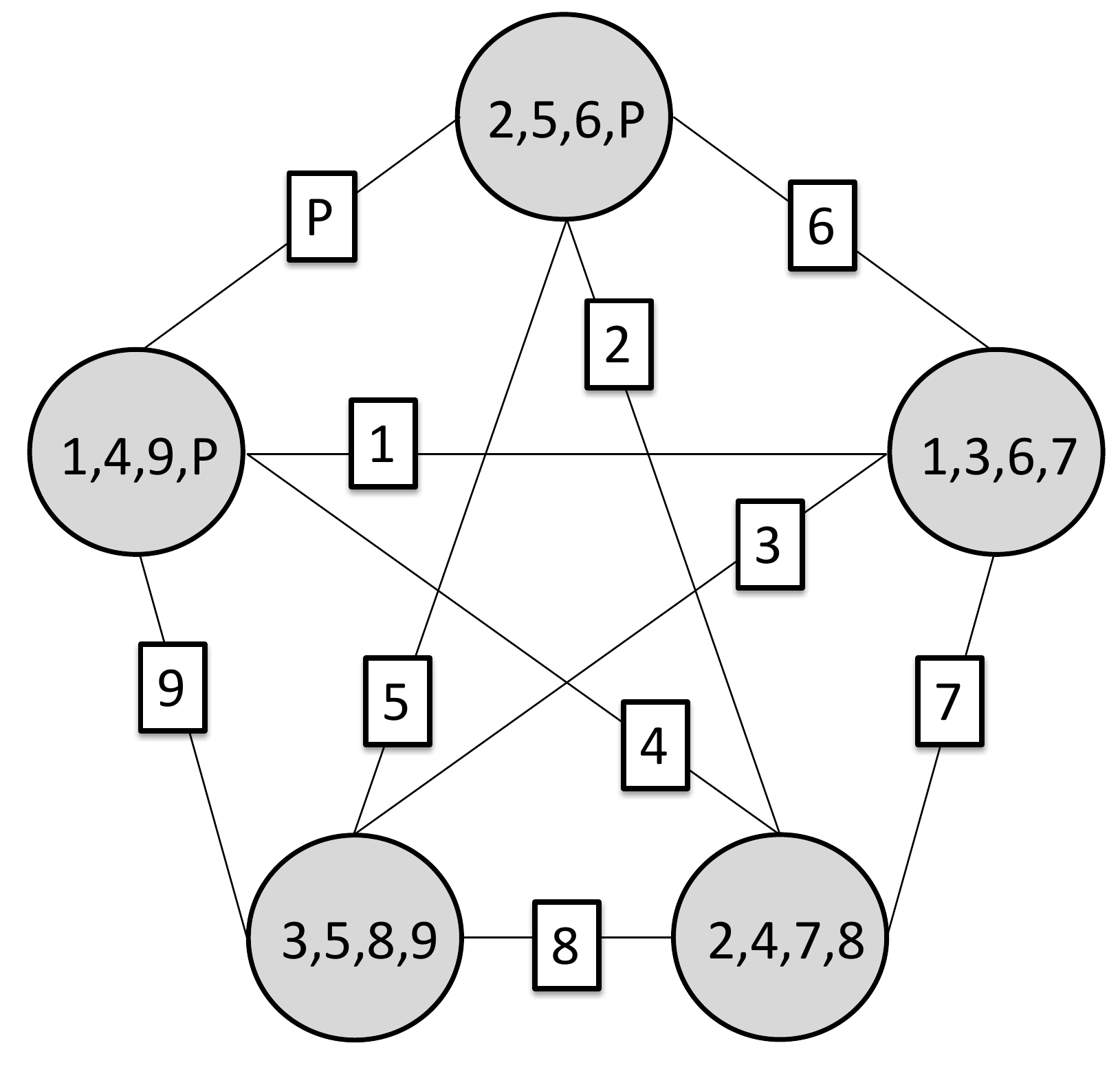}
				\caption{An illustration of pentagon MBR code.}
						\label{fig:pentagon}  
	\end{figure}      
\end{center}

We present a simple construction of an MBR code \cite{RasShaKumRam_allerton09} possessing the RBT property, through an example known as the Pentagon MBR code. The parameters of the example construction are 
\begin{eqnarray*}
	\{(n=5,k=3,d=4),(\alpha=4,\beta=1), B=9 ,\mathbb{F}_2\}. 
\end{eqnarray*}
Note that as required of an MBR code, 
\begin{eqnarray*}
	B = kd\beta - {k \choose 2}\beta=9, & \text{and} & \alpha=d\beta=4.
\end{eqnarray*}
The file to be stored consists of the $9$ symbols $\{a_1,a_2,\cdots,a_9\}$. We first generate a parity symbol $a_P$ given by $a_P=a_1+a_2+\cdots+a_{9} \pmod{2}$. Next, set up a complete graph with $n=5$ nodes, i.e., form a fully-connected pentagon. The pentagon has ${5 \choose 2}=10$ edges, and we assign each of the $10$ code symbols $\{a_i \mid 1 \leq i \leq 9\} \cup \{a_P\}$ to a distinct edge (see Fig.~\ref{fig:pentagon}). Each node stores all the symbols appearing on edges incident on that node, giving $\alpha=4$ .	

{\em Data Collection:} The data collection property requires that the entire data file  be recoverable by connecting to any $k=3$ nodes. It can be easily seen that any collection of $3$ nodes contain $9$ distinct symbols from  $\{a_i \mid 1 \leq i \leq 9\} \cup \{a_P\}$, which is sufficient to recover the entire file.

{\em Node Repair:} Now suppose one of the nodes failed. The repair is accomplished by downloading from  each of the remaining $d=4$ nodes, the code symbol it has in common with the failed node.  Since each helper node passes on a single symbol to aid in node repair, we have $\beta=1$.

This construction can be generalized by replacing the pentagon by a polygon with $n$ vertices.   The parameters of the general construction are
\begin{eqnarray*}
	\{(n,k,d=n-1), \ (\alpha=n-1,\beta=1), \ B= k(n-1)-{k \choose 2},\ \mathbb{F}_q\},  
\end{eqnarray*}
where $q=O(n^2)$. In the initial step of this construction, an $\Large[{n \choose 2},B]\Large$ MDS code is used to generate the code symbols and these code symbols are then assigned to distinct edges of a complete graph on $n$ vertices. Each node stores the symbols appearing on incident edges, resulting in $\alpha=n-1$.  The data collection and node-repair properties in the case of this general construction are easily verified. 

\subsection{The Product-Matrix MSR and MBR Constructions}

The Product-Matrix (PM) framework introduced in \cite{RasShaKum_pm11} allows the construction of both MBR and MSR codes.  For codes constructed under this framework, the parameter $\beta$ is the smallest value possible, i.e., is always equal to $1$. The framework allows the construction of MBR code with $\alpha=d$ for all parameter sets $(n,k,d)$.   In the MSR case, the framework yields constructions for all $(n,k,d)$ with  $\beta=1$ and $d\ge 2k-2$. 

The mathematical setting under this framework is as follows. There is  an $(n \times \alpha)$ code matrix  $C$ whose $i$-th row, denoted by $c_i^t$, contains the $\alpha$ symbols stored in node $i$. The code matrix $C$ is the product of an $(n \times d)$ encoding matrix  $\Psi$ and a $(d\times \alpha)$ message matrix $M$, i.e., $C=\Psi M$. 
The entries of the encoding matrix $\Psi$ are independent of the file to be stored. The message matrix  $M$ contains the $B$ symbols in the file, with some symbols repeated.  Let $\psi_i^t$ denote the $i$-th row of $\Psi$, then the content of node $i$ is given by $c_i^t=\psi_i^tM$.

\subsubsection{PM-MSR Code}

We will begin by identifying the encoding and message matrices that will yield an MSR code with $\beta=1$ \cite{RasShaKum_pm11} for the case $d=2k-2$. This construction can be extended to  $d>2k-2$ through the mechanism of shortening a code \cite{RasShaKum_pm11}. When $d=2k-2$, we have $\alpha=k-1$, $d=2\alpha$ and $B=\alpha(\alpha+1)$.  
The encoding matrix $\Psi$ is then given by $\Psi = [\Phi~\Lambda \Phi]$, where $\Phi$ is a $(n \times \alpha)$ matrix and $\Lambda$ is a $(n \times n)$ diagonal matrix. The entires of $\Psi$ are chosen such that any $d$ rows of $\Psi$ are linearly independent, any $\alpha$ rows of $\Phi$ are linearly independent and the diagonal entries of $\Lambda$ are distinct. These conditions can be meet by picking $\Psi$ to be a Vandermonde matrix that is of the form $\Psi = [\Phi~\Lambda \Phi]$ (this is not difficult). 

Let $S_1$ and $S_2$ be two distinct $(\alpha \times \alpha)$ symmetric matrices, which together contain all the $B=\alpha(\alpha+1)$ elements contained in the data file.  The message matrix $M$ is then given by: $M = \left[ \begin{array}{cc}
S_1 & S_2
\end{array}\right]^t$. Let $\phi_i^t$ denote the $i$-th row of $\Phi$ and $\lambda_i$, the $i$-th diagonal element of the diagonal matrix $\Lambda$. Then the $\alpha$ symbols stored in node $i$ are given by: $c_i^t = \psi_i^t M = \phi_i^t S_1 + \lambda_i \phi_i^t S_2$. We will now show that the data collection and node repair properties hold for this construction.

{\em Data Collection:} Let $\Psi_{\text{DC}} = [\Phi_{\text{DC}}~ \Lambda_{\text{DC}} \Phi_{\text{DC}}]$ be the $(k \times d)$ sub matrix of the $(n \times d)$ matrix $\Psi$ and corresponding to an arbitrary subset of $k$ nodes drawn from the totality of $n$ nodes. To establish the data collection property it suffices to show that one can recover $S_1$ and $S_2$ from the matrix 
\begin{eqnarray*}
	\Psi_{\text{DC}} M = \Phi_{\text{DC}} S_1 + \Lambda_{DC} \Phi_{\text{DC}} S_2, 
\end{eqnarray*}
given $\Phi_{\text{DC}}$ and $\Lambda_{\text{DC}}$.  The first step is to multiply both sides of the equation on the right by the matrix $\Phi_{\text{DC}}^t$ to obtain 
\begin{eqnarray*}
	\Psi_{\text{DC}} M \Phi_{\text{DC}}^t =\Phi_{\text{DC}} S_1 \Phi_{\text{DC}}^t + \Lambda_{DC} \Phi_{\text{DC}} S_2 \Phi_{\text{DC}}^t . 
\end{eqnarray*}
Set $P = \Phi_{\text{DC}} S_1 \Phi_{\text{DC}}^t$ and $Q = \Phi_{\text{DC}} S_2 \Phi_{\text{DC}}^t$, so that $\Psi_{\text{DC}} M \Phi_{\text{DC}}^t =P + \Lambda_{DC} Q $. It can be seen that $P$ and $Q$ are symmetric matrices. The $(i,j)$-th element of $P + \Lambda_{DC} Q$ is $P_{ij}+\lambda_iQ_{ij}$, whereas the $(j,i)$-th element is $P_{ji}+\lambda_jQ_{ji}=P_{ij}+\lambda_jQ_{ij}$. Since all the $\{\lambda_i\}$ are distinct, one can solve for $P_{ij}$ and $Q_{ij}$ for all $i \ne j$, thus obtaining all the non-diagonal entries of both matrices $P$ and $Q$. 
The $i$-th row of $P$ excluding the diagonal element is given by
$\phi_i^t S_1 [\phi_1 \cdots \phi_{i-1}~\phi_{i+1} \cdots \phi_{{\alpha}+1}]$, from which the vector $\phi_i^t S_1$ can be obtained, since the matrix on the right is invertible.   Next, one can form $\left[ 
\phi_1  \cdots  \phi_{\alpha} \right]^t S_1$.  Since the matrix on the left of $S_1$ is invertible, we can then recover $S_1$.  In a similar manner, the entries of the matrix $S_2$ can be recovered from the non-diagonal entries of $Q$.

{\em Node Repair:} Let $f$ be the index of the failed node and $\{h_j | j =1, \dots, d\}$ denote the arbitrary set of $d$ helper nodes chosen. The helper node $h_i$ computes $\psi_{h_i}^t M \phi_f$ and passes it on to the replacement node. Set $\Psi_{rep}=[\psi_{h_1}~\psi_{h_2}\cdots \psi_{h_d}]^t$. Then the $d$ symbols obtained by the replacement node from the repair node can be aggregated into the form $\Psi_{\text{rep}}M \phi_f$. From the properties of $\Psi$, it can be seen that $\Psi_{\text{rep}}$ is invertible. Thus the replacement node can recover $M\phi_f= [S_1\phi_f~S_2\phi_f]^t$. Since $S_1$ and $S_2$ are symmetric matrices, $\phi_f^tS_1$ and $\phi_f^tS_2$ can be obtained by simply taking the transpose. Now $\phi_f^tS_1+ \lambda_f\phi_f^tS_2=\phi_f^tM=c_f^t$, completing the repair process. 

\subsubsection{PM-MBR Code}

For the sake of brevity, we describe here only the structure of the encoding and message matrices under the product-matrix framework, that will result in an MBR code with $\beta=1$.  A proof of the data collection and node-repair properties can be found in \cite{RasShaKum_pm11}.  

From the properties described earlier in Section~\ref{subsec:msr_mbr} of an MBR code it follows that $\alpha=d\beta=d$ and $B=kd-{k \choose 2}$. The $(n \times d)$ encoding matrix $\Psi$ takes on the form $\Psi\  = \left[\Phi ~ \Delta  \right]$, where $\Phi$ is an $(n \times k)$ matrix and $\Delta$ is an $(n \times (d-k))$ matrix.  The matrices are chosen such that any $d$ rows of $\Psi$ are linearly independent and any $k$ rows of $\Phi$ are linearly independent. We remark that these requirements can be satisfied by choosing $\Psi$ to be a Vandermonde matrix.  This places an $O(n)$ requirement on the field size. 
The number of message symbols $B=kd-{k\choose 2}$ can be expressed in the form $B={k+1 \choose 2}+k(d-k)$.  Accordingly, let the $B$ message symbols be divided into two subsets $A_1,A_2$ of respective sizes ${k+1 \choose 2}$ and $k(d-k)$. Let $S$ be a $(k\times k)$ symmetric matrix whose distinct entries correspond precisely to the set $A_1$ (placed in any order).  The symbols in $A_2$ are used to fill up, again in any order, a $(k \times (d-k))$ matrix $T$.  

Given the matrices $S,T$, the $(d\times d)$ symmetric message matrix $M$ is then formed by setting:
\begin{eqnarray*}
	M\  = \left[ \begin{array}{cc}
		S & T \\ 
		T^t & 0
	\end{array} \right].
\end{eqnarray*}
For the repair of failed node $i$, the $j$th helper node sends  $\psi_j^tM\psi_i$. 

\subsection{Clay Code}

\begin{figure}[ht!]
	\begin{center}
		\includegraphics[width=0.5\textwidth]{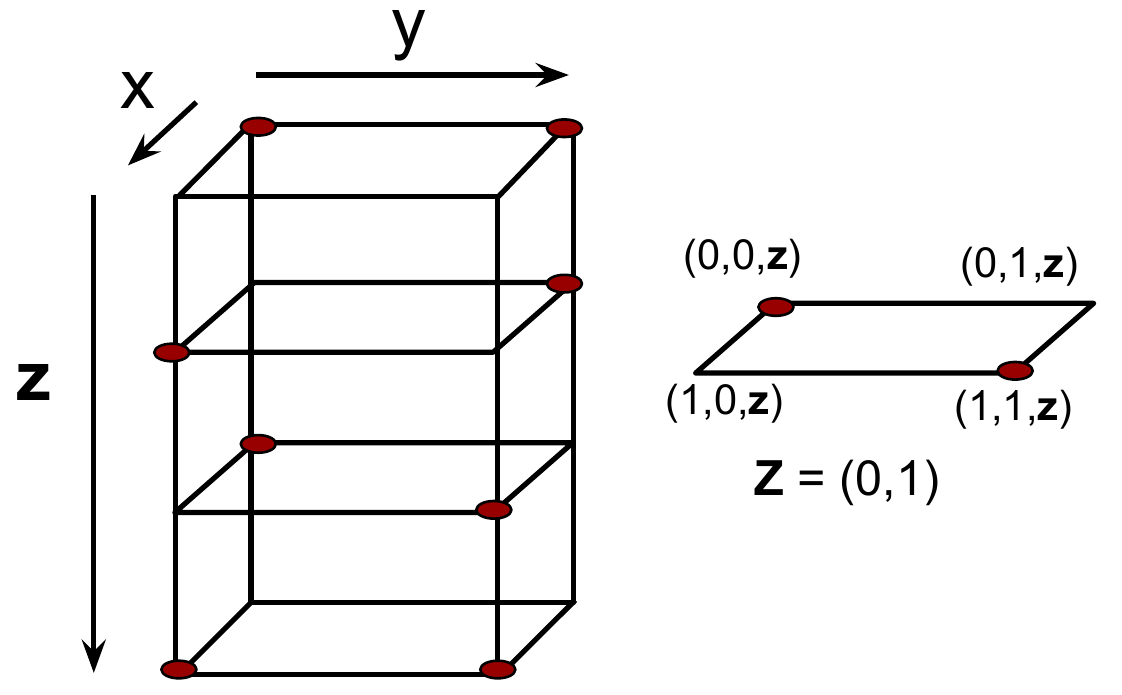}
	\caption{Illustrating the data cube associated with a Clay code having parameters $(r=2,t=2)$. Hence $n=rt=4$ and $\alpha=r^t=4$.  We associate with each codeword in this example Clay code, a $(2 \times 2 \times 4)$ data cube.  Each of the $16$ points within the data cube is thus associated with a distinct code symbol of the codeword.  The data cube is made up of $r^t=4$ planes and each plane is represented by a vector $\mathbf{z}$.  The vector $\mathbf{z}$ associated with a plane in the cube, is identified by the location of dots within the plane, which are placed at the coordinates $(x, y; \mathbf{z})$ satisfying $x = z_y$. \label{fig:datacube} }	
\end{center}
\end{figure}

In this subsection, we present the construction of an optical-access MSR code with minimum-possible level of sub-packetization, having a coupled-layer structure and which is therefore sometimes referred to as the Clay code.  The Clay code was first introduced by Ye-Barg~\cite{YeBar_2} and then independently discovered in \cite{SasVajKum_arxiv}. Work that is very closely related to the Clay code can be found in \cite{LiTangTian}. A systems implementation and evaluation of the Clay code appears in \cite{VajRamPur_Clay}. We use the notation $[a:b] = \{a, a+1, \ldots, b\}$. Clay codes are MSR codes that possess the optimal-access property, have optimal sub-packetization level and can be constructed over a finite field $\mathbb{F}_q$ of any size $q \ge n$.   The parameters of a Clay code construction are of the form: 
\begin{eqnarray*}
	\{(n=rt, k=r(t-1), d=n-1), \ (\alpha=r^t, \beta = r^{t-1}), \ B = k \alpha, \ \mathbb{F}_q \}, 
\end{eqnarray*}
where $q \ge n$, for some $t >1$ and $r \ge 1$.  Each codeword in the Clay code is comprised of a total of $n\alpha= (r \times t \times r^t)$ symbols over the finite field $\mathbb{F}_q$.  We will refer to these finite field symbols as code symbols of the codeword.  These $n \alpha$ code symbols will be indexed by the three tuple 
\begin{eqnarray*}
	(x, y; \mathbf{z}) \ \text{ where } \ x \in [0:r-1], \ y \in [0:t-1], \text{ and } \mathbf{z} = z_0z_1\cdots z_{t-1} \in \mathbb{Z}_r^t.
\end{eqnarray*}
Such an indexing allows us to identify each code symbol with an interior or exterior point of an ($r \times t \times r^t$) three-dimensional (3D) cube and an example is shown in Fig.~\ref{fig:datacube}. For a code symbol $A(x, y; \mathbf{z})$ indexed by $(x, y; \mathbf{z})$, the pair $(x,y)$ indicates the node to which the code symbol belongs, while $\mathbf{z}$ is used to uniquely identify the specific code symbol within the set of $\alpha=r^t$ code symbols stored in that node. 

\textbf{Uncoupled code} As an intermediate step in describing the construction of the Clay code $\mathcal{A}$, we introduce a second code $\mathcal{B}$ that has the same length $n=rt$, the same level $\alpha=r^t$ of sub-packetization and which possesses a simpler description.  For reasons that will shortly become clearer, we shall refer to the code $\mathcal{B}$ as the uncoupled code.  The uncoupled code $\mathcal{B}$ is simply described in terms of a set of $r \alpha$ parity-check (p-c) equations. Let $\{B(x,y;\mathbf{z}) \mid x \in [0:r-1], y \in [0:t-1], \mathbf{z} \in \mathbb{Z}_r^t \}$ be the $n \alpha$ code symbols corresponding to code $\mathcal{B}$. 

The $r \alpha$ p-c equations satisfied by the code symbols that make up each codeword in code $\mathcal{B}$ are given by:
\begin{eqnarray}
\label{eq:uncoup_pc}\sum\limits_{x=0}^{r-1}\sum\limits_{y=0}^{t-1} \theta_{x,y}^{\ell} B(x, y; \mathbf{z}) &=& 0 \text{ for all } \ell \in [0:r-1], \ \ \mathbf{z} \in \mathbb{Z}_r^t,
\end{eqnarray}
where the $\{\theta_{x,y}\}$ are all distinct. Such a $\{\theta_{x,y}\}$ assignment can always be carried out with a field of size $q \ge n = rt$.  The uncoupled code is also an MDS code as it is formed by simply stacking $\alpha$ codewords, each belonging to the same $[n, k]_q$ MDS code.  \\

%

\begin{figure}[ht!]
	\begin{center}
		\includegraphics[width=0.5\textwidth]{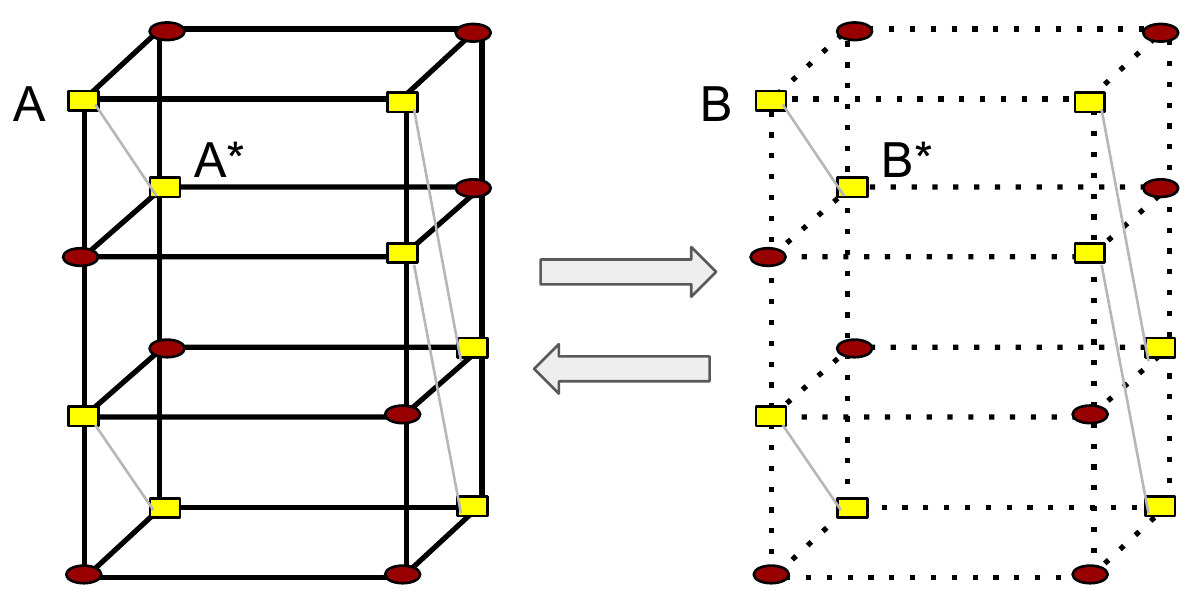}
			\caption{Figure illustrating the data cube with Clay code symbols on the left and the Uncoupled code symbols on the right. $A, A^*$ and $B, B^*$ indicate the paired symbols in the respective cubes.}
						\label{fig:pairing}
	\end{center}
\end{figure}

\textbf{Pairwise Coupling} Next, we introduce a pairing among the $n\alpha$ code symbols (see Fig:\ref{fig:pairing}) associated with uncoded codeword $B(x,y; \mathbf{z})$.  The pair of a code symbol $B(x,y; \mathbf{z})$, for the case $x \ne z_y $ is given by $B(z_y, y; \mathbf{z}(x \rightarrow z_y))$ where the notation $\mathbf{z}(x \rightarrow z_y)$ denotes the vector $\mathbf{z}$ with the $y$th component $z_y$, replaced by $x$, i.e.:
\begin{eqnarray*}
	\mathbf{z}(x \rightarrow z_y) &=&  z_0 z_1 \cdots z_{y-1} x z_{y+1} \cdots z_{t-1} \in \mathbb{Z}_r^t.
\end{eqnarray*}
The code symbols $B(x,y; \mathbf{z})$, for the case $x = z_y$ will remain unpaired.  One can alternately view this subset of code symbols as being paired with themselves, i.e., the pair of $B(x,y; \mathbf{z})$, for the case $x = z_y$, is $B(x,y; \mathbf{z})$ itself. 

We now introduce a pairwise transformation, which we will refer to as the coupling transformation, which will lead from a codeword \begin{eqnarray*}
	\left( B(x,y; \mathbf{z}) \mid x \in [0:r-1], y \in [0:t-1], \mathbf{z} \in \mathbb{Z}_r^t \right), 
\end{eqnarray*}
in the coupled code to a codeword 
\begin{eqnarray*}
	\left( A(x,y; \mathbf{z}) \mid x \in [0:r-1], y \in [0:t-1], \mathbf{z} \in \mathbb{Z}_r^t \right), 
\end{eqnarray*}
in the coupled code.  For $x \ne z_y$ the pairwise transformation takes on the form:   
\begin{eqnarray}
\label{eq:coup}\left[\begin{array}{c}
A(x, y; \mathbf{z})\\
A(z_y, y, \mathbf{z}(x \rightarrow z_y ))
\end{array}\right] &=& \left[\begin{array}{cc}
1 & \gamma\\
\gamma & 1
\end{array} \right]^{-1} \left[\begin{array}{c}
B(x, y; \mathbf{z})\\
B(z_y, y; \mathbf{z}(x \rightarrow z_y))
\end{array}\right],
\end{eqnarray}
where $\gamma$ is selected such that $\gamma^2\neq 1$.  This causes the $(2 \times 2)$ linear transformation above to be invertible.  For the case $x = z_y$, we simply set 
\begin{eqnarray*}
	A(x, y; \mathbf{z}) = B(x, y; \mathbf{z}).
\end{eqnarray*}


\textbf{Parity Check Equations} Combining equations \eqref{eq:coup} and \eqref{eq:uncoup_pc} gives us the p-c equations satisfied by the code symbols of the Clay Code ${\cal A}$:
\begin{eqnarray}
\label{eq:coup_pc}
\sum\limits_{x = 0}^{r-1}\sum\limits_{y=0}^{t-1} \theta_{x,y}^{\ell} \left( \ A(x, y; \mathbf{z}) + \mathsf{1}_{\{x \ne z_y\}} A(x, y; \mathbf{z}(x \rightarrow z_y)) \ \right) &=& 0,
\end{eqnarray}
for all $\ell \in [0:r-1]$ and all $\mathbf{z} \in \mathbb{Z}_r^t$, where $\mathsf{1}_{\{x \ne z_y\}}$ is equal to $1$ if $x \ne z_y$ or else takes the value $0$.

\begin{figure}[ht!]
	\begin{center}
		\includegraphics[width=0.8\textwidth]{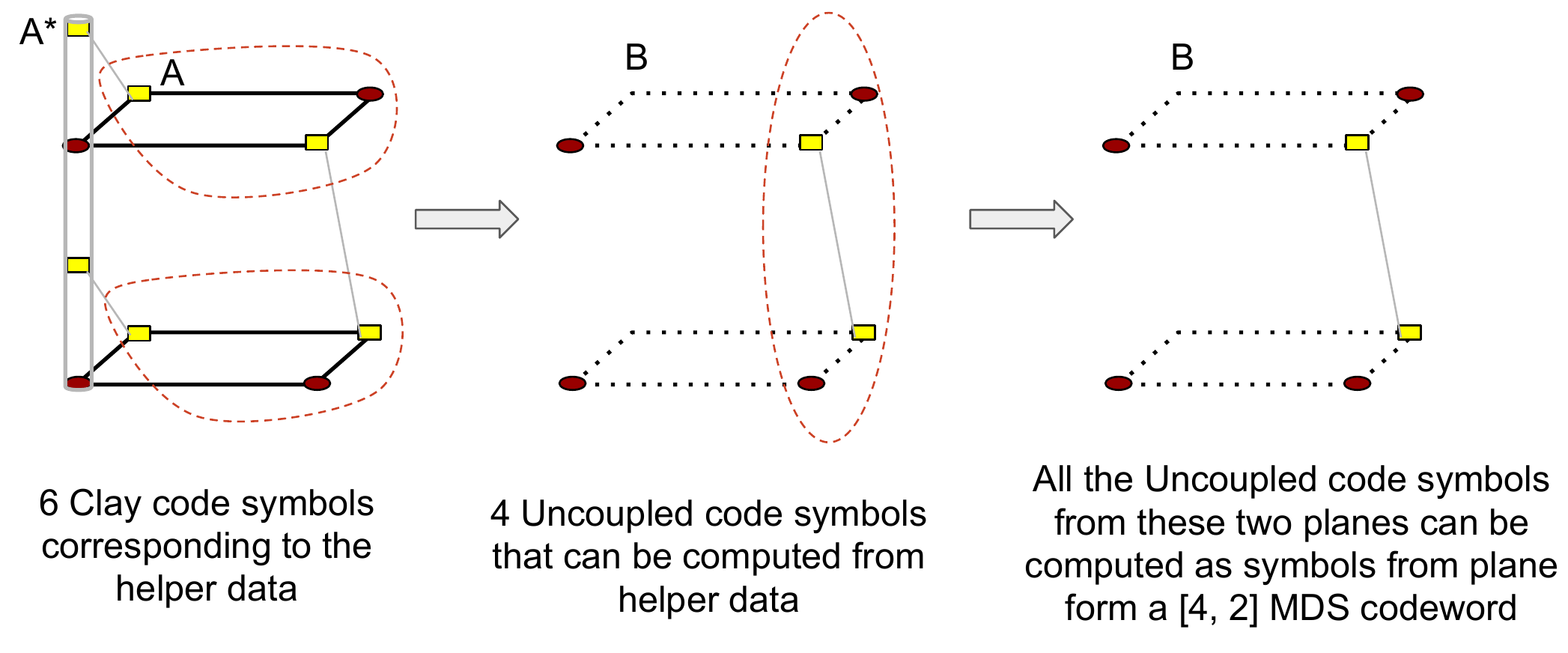}
	\caption{Figure illustrating node repair of $(x_0, y_0) = (1,0)$. The helper data sent corresponds to repair planes $\mathbf{z}$ with a dot at failed node $(1,0)$. The Uncoupled code symbols corresponding to repair planes are recovered as shown above. We can therefore recover the failed symbols from the repair planes and also the remaining failed node symbols of the form $A^*$ from $A, B$ as shown in the figure.} 
		\label{fig:noderepair}	
\end{center}
\end{figure}

\textbf{Optimal-Access Node Repair} We will show how repair of single node $(x_0, y_0)$ in the Clay code is accomplished by downloading $\beta$ symbols from each of the remaining $n-1$ nodes.  Since no computation is required at a helper node, this will also establish that the Clay code possesses the optimal-access property.  The $\beta = r^{t-1}$ symbols passed on by a helper node $(x,y) \ne (x_0, y_0)$ are precisely the subset $\{A(x, y; \mathbf{z}) \mid \mathbf{z} \in P \}$ of the $\alpha=r^t$ symbols contained in that node, in which 
\begin{eqnarray*}
	P = \{\mathbf{z} \in \mathbb{Z}_r^t \mid z_{y_0}=x_0  \},
\end{eqnarray*} 
identifies an $r^{t-1}$-sized subset of the totality of $r^t$ planes in the cube.  We will refer to $P$ as the set of repair planes.  Pictorially, these are precisely the planes that have a dot in the location of the failed node, see Fig.~\ref{fig:noderepair}.\\  
Consider the $r$ p-c equations given by \eqref{eq:coup_pc}for a fixed repair plane $\mathbf{z}$, i..e, a plane $\mathbf{z}$ belonging to $P$.   The code symbols appearing in these equations either belong to the plane $\mathbf{z}$ itself, or else are paired to a code symbol belonging to $\mathbf{z}$.   For $y \neq y_0$, the pair of a code symbol $A(x,y,\mathbf{z})$ lying in $\mathbf{z}$, is a second code symbol $A(z_y,y,\mathbf{z}(x \rightarrow z_y))$ that does not belong to $\mathbf{z}$, but lies however, in a different plane $\hat{\mathbf{z}}$ that is also a member of $P$.  It follows that for $y \neq y_0$, the replacement for failed node $(x_0,y_0)$, has access to all the code symbols with $y \neq y_0$ that appear in \eqref{eq:coup_pc}.  For the case $y=y_0$, $x \ne x_0$ the replacement node has access to code symbols that belong to the plane $\mathbf{z}$, but not to their pairs.  Based on the above, the p-c equations corresponding to the plane $\mathbf{z} \in P$ can be expressed in the form: 
\begin{eqnarray*}
	\theta_{x_0, y_0}^{\ell} A(x_0, y_0; \mathbf{z}) + \sum\limits_{x \ne z_{y_0}} \gamma \theta_{x, y_0}^{\ell} A(x_0, y_0; \mathbf{z}(x \rightarrow z_{y_0})) &=& \kappa^*,
\end{eqnarray*}
where $\kappa^*$ indicates a quantity that can be computed at a replacement node, based on the code symbols supplied by the $(n-1)$ helper nodes.   
Thus we have a set of $r$ equations in $r$ unknowns.  From the theory of generalized RS codes, it is known that these equations suffice to recover the symbols: 
\begin{eqnarray*}
	\{A(x_0, y_0; \mathbf{z}(x \rightarrow z_{y_0})) \mid  \mathbf{z} \in P, x \in [0:r-1] \} = \{A(x_0, y_0, \mathbf{z} ) \mid \mathbf{z} \in \mathbb{Z}_r^t\}.
\end{eqnarray*}
This completes the proof of node repair.  We refer the readers to \cite{SasVajKum_arxiv} for a proof of the data collection property.  

\subsection{Variants of Regenerating Code}

\textbf{Cooperative RGC} Regenerating codes were introduced initially with the aim of efficiently dealing with single node-failures, subsequent work extended their applicability to the simultaneous failure of $t > 1$ nodes.  Two approaches were adopted. Under the first approach, there is a single repair center that collects all the helper information pertinent to the repair of all the failed nodes and then subsequently distributes the requisite information to the replacement nodes. In \cite{CadJafMalRamSuh}, the notion of interference alignment is applied to establish that the total amount of data downloaded to the repair center for the case when the code is an MDS code, is at least $\frac{\alpha dt}{d+t-k}$.

Under the second approach, there are $t$ repair centers, each corresponding to a replacement node.   In the first phase of the repair process, helper data is transferred to the $t$ repair centers from the $d\leq (n-t)$ helper nodes.  In the second phase, data is transferred between repair centers.  This latter approach introduced in \cite{HuXuWanZhaLi} is known as cooperative regeneration. A storage-repair-bandwidth tradeoff for cooperative regenerating codes was derived in \cite{KerScoStr} and \cite{ShuHu} using cutset-bound-based techniques. The extreme points of the tradeoff are known as minimum-bandwidth cooperative regenerating (MBCR) code and minimum-storage  cooperative regenerating (MSCR) points. Constructions of optimal codes that operating at these extreme points can be found in \cite{WanZhaCoopRegen} and \cite{YeBarg_Coop}. 

\textbf{Near Optimal Bandwidth MDS Codes} Yet another variant of RGCs explored in the literature are vector MDS codes that trade between sub-packetization level and savings in repair bandwidth. As demonstrated in \cite{VajRamPur_Clay}, large sub-packetization level is not a desirable feature in a distributed storage system. Though MSR codes have optimal repair bandwidth, they necessarily incur a high sub-packetization level as established in \cite{TamWanBru_access_tit,GopTamCal,BalKum_subpkt,AlrabGur}. The piggybacking framework \cite{RasShaRam_Piggyback}, $\epsilon-$MSR framework \cite{RawTamGur_epsilonMSR} and the transformation in \cite{LiTang} are examples of construction methods for vector MDS codes that have a small sub-packetization level while ensuring substantial reduction in repair bandwidth. 


\textbf{Fractional-Repetition Codes}, introduced in \cite{RouRamFR10} may be regarded as a generalization of the RBT polygon MBR code \cite{RasShaKumRam_allerton09} presented in Section~\ref{sec:Polygon_MBR}. In a fractional-repetition code, the symbols corresponding to a data file are first encoded using an MDS code. Each code symbol is then replicated $\rho$ times. These replicated code symbols are stored across $n$ nodes, with each node storing $\alpha$ symbols and each code symbol appearing in exactly $\rho$ distinct nodes.  The definition of MBR codes requires that any set of $d$ surviving nodes can be called upon to serve as helper nodes for node repair.  In contrast, in the case of fractional-repetition codes, it is only guaranteed that there is at least one choice of $d$ helper nodes corresponding to which RBT is possible. A fractional-repetition code with $\rho$-repetition allows repair without any computation, for up to $\rho-1$ node failures.  Constructions of fractional-repetition codes can be found in \cite{RouRamFR10, PawNoorRouRamDressCode11,SilbEtzOptimalFR15,OlmRamFRCodes16}.

\textbf{Secure RGCs} introduced in \cite{PawarRouayRam}, are a variant of RGCs 
where a passive but curious eavesdropper is present, who has access to the data stored in a subset $A$ of size $|A|=\ell < k$ of the storage nodes.
The aim here is to prevent the eavesdropper from gaining any information pertinent to the stored data. Here again, there is a notion of exact and functional repair and there are corresponding storage-repair bandwidth tradeoffs.  Codes that operate at extreme ends of the tradeoff are called secure MBR and secure MSR codes respectively.  In \cite{PawarRouayRam} an upper bound on file size for the case of F-R secure RGC is provided along with a matching construction corresponding to the MBR point for the case $d=n-1$. In \cite{ShaoLiuTianShen}, the authors study the E-R storage-bandwidth tradeoff for this model.

This eavesdropper model was subsequently extended in \cite{ShahRashKum} to the case where the passive eavesdropper also has access to data received during the repair of a subset of nodes $A_1 \subseteq A$ of size $\ell_1$.  In \cite{ShahRashKum}, the authors provide a secure MBR construction that holds for any value of parameter $d$.  The file size under this construction, matches the upper bound shown in \cite{PawarRouayRam}. In \cite{Rawat_secrecy}, for the extended model, the authors provide an upper bound on file size corresponding to the secure MSR point and a matching construction. In \cite{YeShumYeung}, the authors study the E-R storage-bandwidth tradeoff for the extended model.

\section{Locally Recoverable Codes} 
In the case of the class of regenerating codes discussed in the previous section, the aim was to reduce the repair bandwidth.  In contrast, the focus in the case of Locally Recoverable Codes (LRCs) introduced in \cite{HanMon,HuaCheLi} and discussed in the present section, is on reducing the repair degree, i.e., on reducing the number of helper nodes contacted for the purpose of node repair. Two comments are in order here. Firstly, reducing the repair degree does tend to lower the repair bandwidth.  Secondly, the storage overhead in the case of a non-trivial LRC, is necessarily larger than that of an MDS code.  There are two broad classes of LRCs. LRCs with \textbf{Information Symbol Locality} (ISL) are systematic linear codes in which the repair degree is reduced only for the repair of nodes corresponding to message symbols.  In an LRC with \textbf{All-Symbol Locality} (ASL), the repair degree is reduced for the repair of all $n$ nodes, regardless of whether the node corresponds to message or parity symbol.  Clearly, the class of LRCs with ASL is a sub-class of the set of all LRCs with ISL. 

\subsection{Information Symbol Locality}
Unless otherwise specified, when we speak on an LRC in this section, the reference will be to an LRC with ISL.  A linear code is said to be systematic if the $k$ message symbols are explicitly present among the $n$ code symbols.  An $(n,k,r)$ LRC $\mathcal{C}$ over a field $\mathbb{F}_q$, is a systematic $[n,k]$ linear block code having the property that every message symbol $c_t$, $t \in [k]$ can be recovered by computing a linear combination of the form 
\begin{eqnarray*}
	c_t & = & \sum_{j \in S_t} a_j c_j , \ \ a_j \in \mathbb{F}_q \ 
\end{eqnarray*} 
involving at most $r$ other code symbols $c_j, j \in S_t$. Thus the set $S_t$ in the equation above has size at most $r$. 
The minimum distance of an $(n,k,r)$ LRC \cite{GopHuaSimYek} must necessarily satisfy the bound
\begin{eqnarray}
d_{\min} & \leq & (n-k+1) - \left( \left \lceil \frac{k}{r} \right \rceil -1 \right). \label{eq:optimaldmin}
\end{eqnarray}
Thus for the same values of $[n,k]$, an LRC has $d_{\min}$ which is smaller by an amount equal to $\left( \left \lceil \frac{k}{r} \right \rceil -1 \right)$ in comparison with an MDS code.  The quantity, $\left( \left \lceil \frac{k}{r} \right \rceil -1 \right)$ may thus be regarded as the penalty associated with imposing the locality requirement.   An LRC whose minimum distance satisfies the above bound with equality is said to be optimal. The class of pyramid codes \cite{HuaCheLi} are an example of a class of optimal LRCs and are described below.  Analysis of non-linear LRCs can be found in \cite{PapDim,ForYekh}.

\subsubsection{Pyramid Codes} \label{pyramid}
We introduce the pyramid code \cite{HuaCheLi} construction of an LRC with ISL through an illustrative example corresponding to parameter set $(n=9,k=6,r=3)$.   The starting point is the generator matrix of an RS code.  Let $G_{RS}$ be the generator matrix of an $[n_{RS}=8,k_{RS}=6]$ RS code $\mathcal{C}_{RS}$ in systematic form, i.e.,  
\begin{eqnarray*}
	G_{RS} = \begin{bmatrix}
		1 & 0 & 0 & 0 & 0 & 0 & \vline & g_{11} & g_{12} \\
		0 & 1 & 0 & 0 & 0 & 0 & \vline & g_{21} & g_{22} \\
		0 & 0 & 1 & 0 & 0 & 0 & \vline & g_{31} & g_{32} \\
		0 & 0 & 0 & 1 & 0 & 0 & \vline & g_{41} & g_{42} \\ 
		0 & 0 & 0 & 0 & 1 & 0 & \vline & g_{51} & g_{52} \\
		0 & 0 & 0 & 0 & 0 & 1 & \vline & g_{61} & g_{62} 
	\end{bmatrix}.
\end{eqnarray*}
The generator matrix of the associated pyramid code is obtained by splitting a single parity column in $G_{RS}$ and then rearranging columns as shown below:   
\begin{eqnarray*}
	G_{\text{pyr}} = \begin{bmatrix}
		1 & 0 & 0 & g_{11} & 0 & 0 & 0 & 0 & \vline & g_{12} \\
		0 & 1 & 0 & g_{21} & 0 & 0 & 0 & 0 & \vline & g_{22} \\
		0 & 0 & 1 & g_{31} & 0 & 0 & 0 & 0 & \vline & g_{32} \\
		0 & 0 & 0 & 0 & 1 & 0 & 0  & g_{41} & \vline & g_{42} \\ 
		0 & 0 & 0 & 0 & 0 & 1 & 0  & g_{51} & \vline & g_{52} \\
		0 & 0 & 0 & 0 & 0 & 0 & 1  & g_{61} & \vline & g_{62} 
	\end{bmatrix}.
\end{eqnarray*}
This yields the generator matrix $G_{\text{pyr}}$ of an $(n=9,k=6,r=3)$ optimal LRC code $\mathcal{C}_{\text{pyr}}$.  The proof that the above code is an optimal LRC with ISL is as follows.  It is clear that the code $\mathcal{C}_{\text{pyr}}$ is an LRC and that the minimum distance $d_{\min}$ of the code $\mathcal{C}_{\text{pyr}}$ is at least the minimum distance of the RS code $\mathcal{C}_{RS}$. This follows from the fact that the minimum Hamming distance of a linear code equals its minimum Hamming weight. The minimum distance of the $\mathcal{C}_{RS}$ equals $n_{RS}-k_{RS}+1=8-6+1=3$ from the Singleton bound.  It follows that the minimum distance of the pyramid code is at least $3$.  On the other hand, from \eqref{eq:optimaldmin}, we have that 
\begin{eqnarray*}
	d_{\min} \leq (n-k+1) - \left( \left \lceil \frac{k}{r} \right \rceil -1 \right) = 9-6+1-\left( \left \lceil \frac{6}{3} \right \rceil -1 \right) = 3. 
\end{eqnarray*}
It follows that the code is an optimal LRC.  
In the general case, if we start with an $[n,k]$ RS code and split a single parity column, we will obtain an optimal 
\begin{eqnarray*}
	(n_{\text{pyr}}=n+\lceil k/r \rceil -1,\ k_{\text{pyr}}=k, \ r)
\end{eqnarray*}
pyramid LRC.

\subsubsection{Windows Azure LRC} \label{sec:Azure} 
\begin{figure}[h!]
	\begin{center} 
		\includegraphics[width=5in]{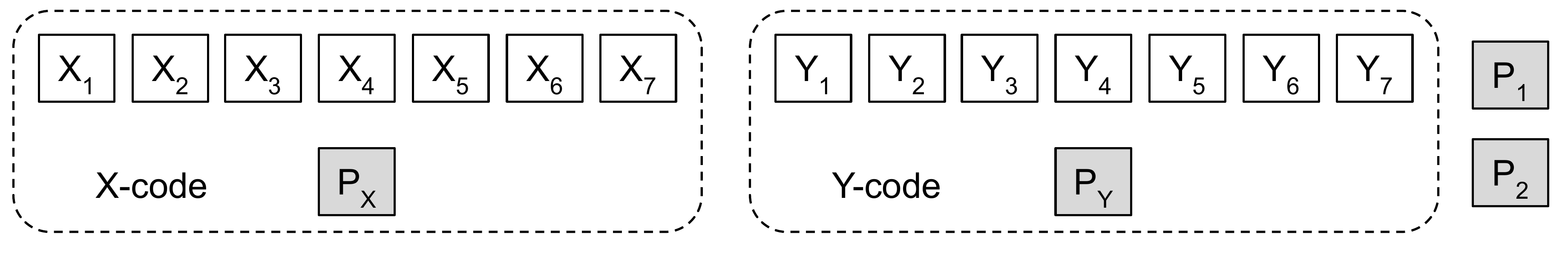}
			\caption{The LRC employed in Windows Azure cloud storage.}  \label{fig:Azure}   
	\end{center}
\end{figure}

Fig.~\ref{fig:Azure} shows the $(n=18,k=14,r=7)$ LRC employed in the Windows Azure cloud storage system \cite{HuaSimXuOguCalGopLiYek} and which is related in structure, to the pyramid code.  The dotted boxes indicate a collection of symbols that satisfy an overall parity check.  This code has minimum distance $4$ which is the same as that of the $[n=9,k=6]$ RS code.

In terms of reliability, the $(n=18,k=14,r=7)$ Windows Azure code and the $[9,6]$ RS code are comparable as they both have the same minimum distance $d_{\min}=4$. In terms of repair degree, the two codes are again comparable, having respective repair degrees of $7$ (Windows Azure LRC) and $6$ (RS).  The major difference is in the storage overhead, which stands at $\frac{18}{14}=1.29$ in the case of the Azure LRC and $\frac{9}{6}=1.5$ in the case of the $[9,6]$ RS code.  This reduction in storage overhead has reportedly saved Microsoft millions of dollars \cite{microsoft}. 

\subsection{All Symbol Locality} \label{TamoBarg}

\begin{figure}[h!]
	\begin{center} 
		\includegraphics[width=2.5in]{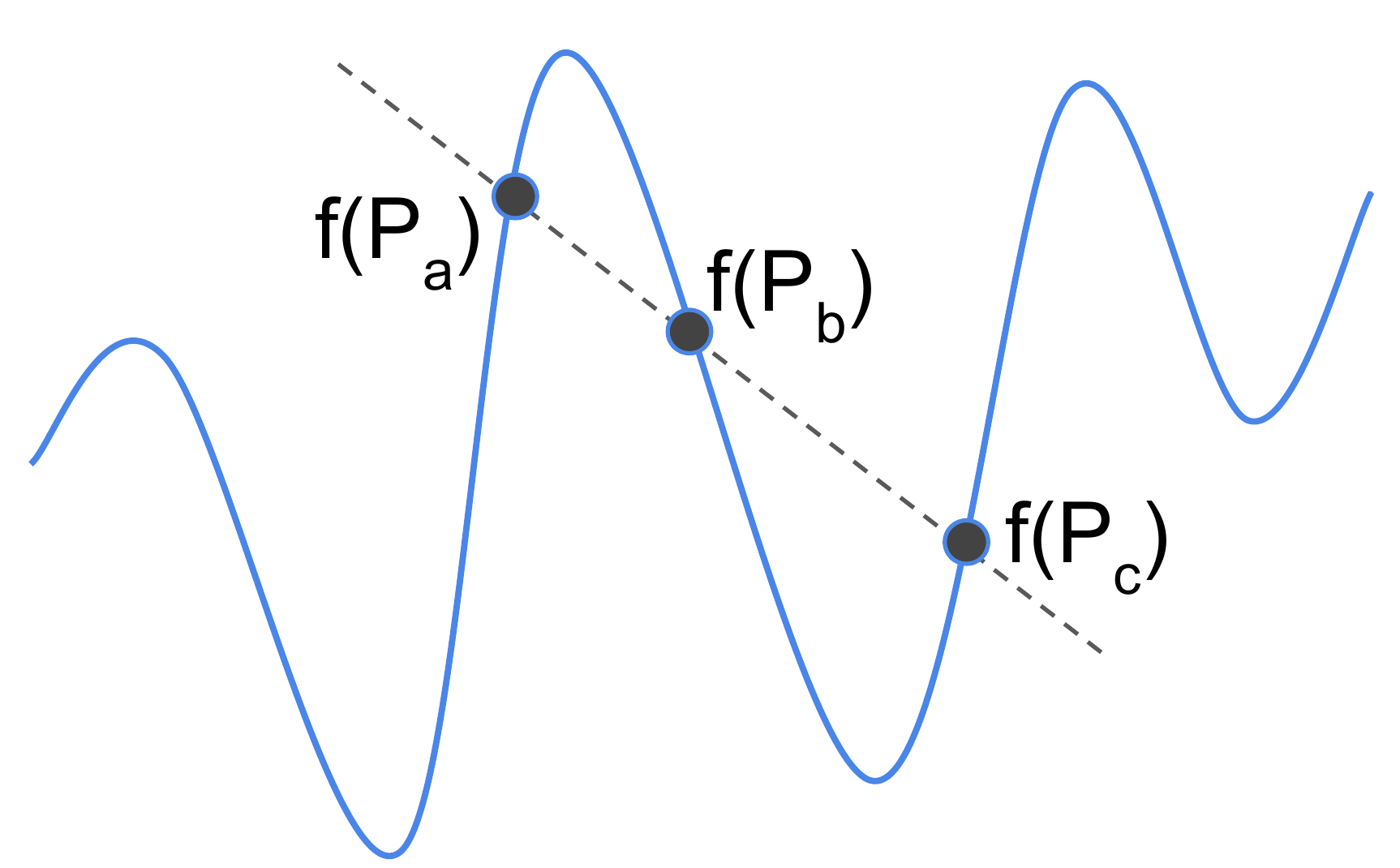}
			\caption{Illustrating the construction of an optimal ASL LRC.}  \label{fig:tamo_barg_lrc}  
	\end{center}  
\end{figure}

An LRC in which every code symbol can be recovered from a linear combination of at most $r$ other code symbols is called an LRC with ASL. An LRC with ASL will be said to be optimal if it has minimum distance that satisfies \eqref{eq:optimaldmin} with equality. A construction for optimal ASL LRCs can be found in \cite{TamBar_LRC}. The codes in the construction may be regarded as subcodes of RS codes. The idea behind the construction is illustrated in Fig.~\ref{fig:tamo_barg_lrc} for the case $r=2$. As noted in Section~\ref{sec:RS}, the code symbols within a  codeword of an $[n.k]$ RS code over $\mathbb{F}_q$ may be regarded as evaluations of a polynomial associated with the message symbols.  More specifically, the codeword $(f(P_1), \cdots f(P_n)) \in \mathbb{F}_q^n$, where $f(x)=\sum_{i=0}^{k-1} m_i x^i$ and where $P_1,\ldots,P_n$ are distinct elements from $\mathbb{F}_q$, is associated to the set $\{m_i\}_{i=0}^{k-1}$ of message symbols.  The construction depicted in Fig.~\ref{fig:tamo_barg_lrc}, is one in which code symbols are obtained by evaluating a subclass of this set of polynomials.  This subclass of polynomials has the property that given any code symbol corresponding to the evaluation $f(P_a)$, there exist two  other code symbols $f(P_b),f(P_c)$ such that the three values lie on straight line and hence satisfy an equation of the form
\begin{eqnarray*}
	u_af(P_a)+ u_bf(P_b)+ u_cf(P_c) & = & 0,   
\end{eqnarray*}
where $u_a,u_b,u_c \subseteq \mathbb{F}_q$.  Thus the value of an example code symbol $f(P_a)$ can be recovered from the values of two other code symbols, $f(P_b)$ and $f(P_c)$ in the present case.  Thus this construction represents an LRC with $r=2$ which structurally is a subcode of an RS code.  

We now present a more general form of the construction in \cite{TamBar_LRC} of an $(n,k,r)$ ASL LRC for the case $n=(q-1)$ and $(r+1)|(q-1)$.  It will be found convenient to express $k$ in the form 
\begin{eqnarray*}
	k & = & \ell r +a, \ \ 1 \leq a \leq r,   
\end{eqnarray*}
so that 
\begin{eqnarray*}
	\ell & = & \left \lceil \frac{k}{r} \right\rceil -1. 
\end{eqnarray*}
While in an RS code, we evaluate all polynomials of degree $\leq (k-1)$, here we restrict attention to the subset $Q$ of polynomials, 
that can be expressed in the form: 
\begin{eqnarray*}
	f(x)=\sum_{i=0}^{\ell} x^{(r+1)i}f_i(x), 
\end{eqnarray*}
where the polynomials 
\begin{eqnarray*}
	f_i(x) & = & \sum_{j=0}^{r-1} b^i_j x^j
\end{eqnarray*}
for $0\leq i\leq \ell-1$ have degree $(r-1)$ and where 
\begin{eqnarray*}
	f_{\ell}(x) & = & \sum_{j=0}^{a-1} b^{\ell}_j x^j
\end{eqnarray*}
has degree $(a-1)$.  Clearly by counting the number of coefficients, we see that the number of polynomials in the set $Q$ equals $q^{\ell r+a}=q^k$ and hence this code has dimension $k$.  Let $\mathbb{F}_q^{*}$ denote the set of $(q-1)$ nonzero elements in the finite field $\mathbb{F}_q$.  Code symbols are obtained by evaluating each polynomial 
in $Q$ at all the elements in $\mathbb{F}_q^{*}$.  Let ${\cal C}$ be the resultant code, i.e., 
\begin{eqnarray*}
	{\cal C} & = & \left\{ (f(u) \mid u \in \mathbb{F}_q^{*}) \mid f \in Q \right\} .
\end{eqnarray*}
We will next establish that ${\cal C}$ is an LRC.  To see this, let $H$ denote the set of $(r+1)^{th}$ roots of unity contained in $\mathbb{F}_q$. Then the $\frac{(q-1)}{(r+1)}$ multiplicative cosets of $H$ partition $\mathbb{F}_q^{*}$. We first note that for any $\beta \in H$ and any $b \in \mathbb{F}_q^{*}$, the product $b\beta$ is a zero of the polynomial $(x^{r+1}-b^{r+1})$.  It follows then that for $f \in Q$, 
\begin{eqnarray}
f(b\beta)&=&f(x)|_{x=b\beta} = \left(f(x) \mod (x^{(r+1)}-b^{r+1}) \right)|_{x=b\beta}, \nonumber \\
&=& \left( \sum_{i=0}^{\ell} x^{(r+1)i}f_i(x) \mod (x^{(r+1)}-b^{r+1}) \right)|_{x=b\beta}, \nonumber  \\
&=&\left( \sum_{i=0}^{\ell} b^{(r+1)i}f_i(x) \right)|_{x=b\beta}. \label{eq:poly_lrc}
\end{eqnarray}
Since each polynomial $f_i(x)$ is of degree $\leq r-1$, the polynomial appearing on the right in equation \eqref{eq:poly_lrc}, is also of degree $\leq r-1$.   As a result, we can recover the value $f(b\beta)$ from the $r$ evaluations $\{ f(b \theta): \theta \in H \setminus \{\beta\} \}$.   Thus this construction results in an LRC with locality parameter $r$.

We will now show that ${\cal C}$ is an optimal LRC with respect to the minimum distance bound in \eqref{eq:optimaldmin}.   We next estimate the minimum distance of the code by computing the maximum degree of a polynomial in $Q$.  
We see that 
\begin{eqnarray*}
	deg(f) &\leq &  \ell (r+1)+a-1 \\
	& = & 	k+ \left( \left\lceil \frac{k}{r} \right\rceil -1 \right) -1.
\end{eqnarray*}
Since a polynomial of degree $d$ can have at most $d$ zeros and the minimum Hamming weight of a linear code equals its minimum distance, it follows that 
\begin{eqnarray}
d_{\min} & \geq & n-\deg(f) \nonumber \\ 
& \geq & n-k+1-\left(\left \lceil \frac{k}{r} \right \rceil-1 \right) .\label{dmintamobarg}
\end{eqnarray}
Comparing \eqref{eq:optimaldmin} and \eqref{dmintamobarg}, we see that the code ${\cal C}$ is an optimal ASL LRC. Note that the field size needed for this construction is $O(n)$.   A different construction of optimal LRC with $O(n)$ field size that is based on cyclic codes can be found in \cite{TambarGopCal}. 

It turns out that for parameter sets where $(r+1) \nmid n$, the bound \eqref{eq:optimaldmin} cannot be achieved with equality by any ASL LRC. Improved versions of the bound \eqref{eq:optimaldmin} can be found in \cite{WanZha_bound,ZhaWanGe,MehArd}. A construction achieving the improved bound in \cite{WanZha} with equality and with exponential field size can be found in the same paper for the case $n_1>n_2$ where $n_1=\lceil \frac{n}{r+1} \rceil$ and $n_2=n_1(r+1)-n$. 

\subsection{LRCs Over Small Field Size}

For field size $q<n$, it is challenging to construct optimal LRCs.  Upper bounds on the minimum distance of an $(n,k,r)$ LRC over $\mathbb{F}_q$ that take into account the field size $q$ can be found in \cite{CadMaz,BalPVK,WanZhaLin,HuaYakUchSie,TambarGopCal,GopCal,AgaBarHuMazTam}.  Example constructions that are optimal with respect to these improved bounds can be found in \cite{WanZhaLin, HuaYakUchSie, GopCal}. 
Asymptotic upper bounds on an LRC i.e., upper bounds on the rate $\frac{k}{n}$ for a fixed value of relative minimum distance $\frac{d_{\min}}{n}$ and in the limit as $n \rightarrow \infty$ and which take into account, the field size $q$, can be found in \cite{AgaBarHuMazTam}.  Asymptotic lower bounds can be found in \cite{CadMaz,BarTamVla}.

\subsection{Recovery from Multiple Erasures} 

\begin{center}
	\begin{figure}[h!]
		\centering
	  	\includegraphics[width=4.2in]{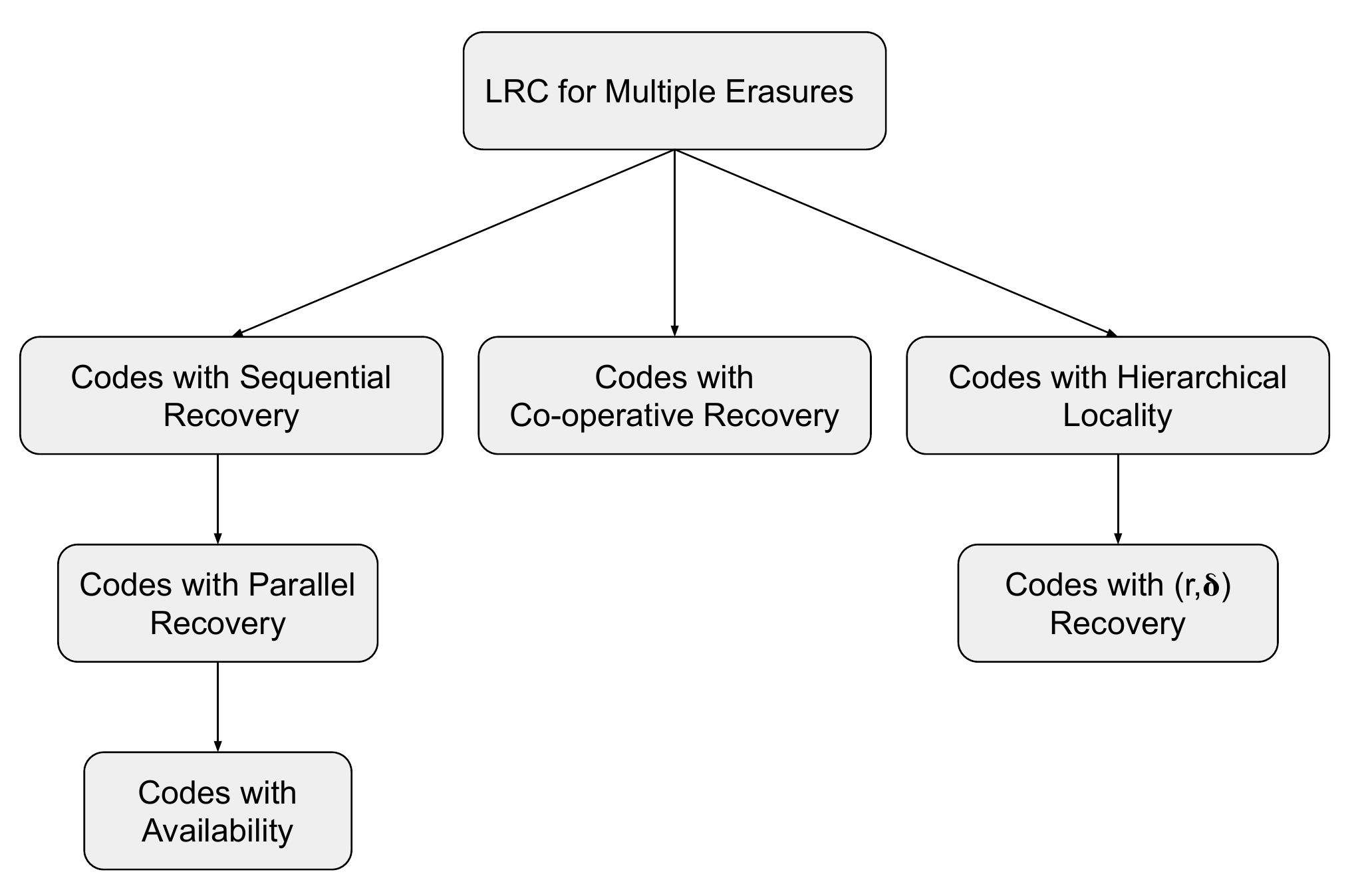}  
	\caption{This figure illustrates the classification of various approaches for LRCs for multiple erasures.}  \label{fig:classifcation_lrc}  	
\end{figure}      
\end{center}

There are several approaches towards designing an LRC that can recover from more than one erasure.  A classification of these approaches is presented in Fig.~\ref{fig:classifcation_lrc}. 
\subsubsection{Codes With Sequential Recovery} 
\begin{center}
	\begin{figure}[h!]
		\centering 
		\includegraphics[width=2.6in]{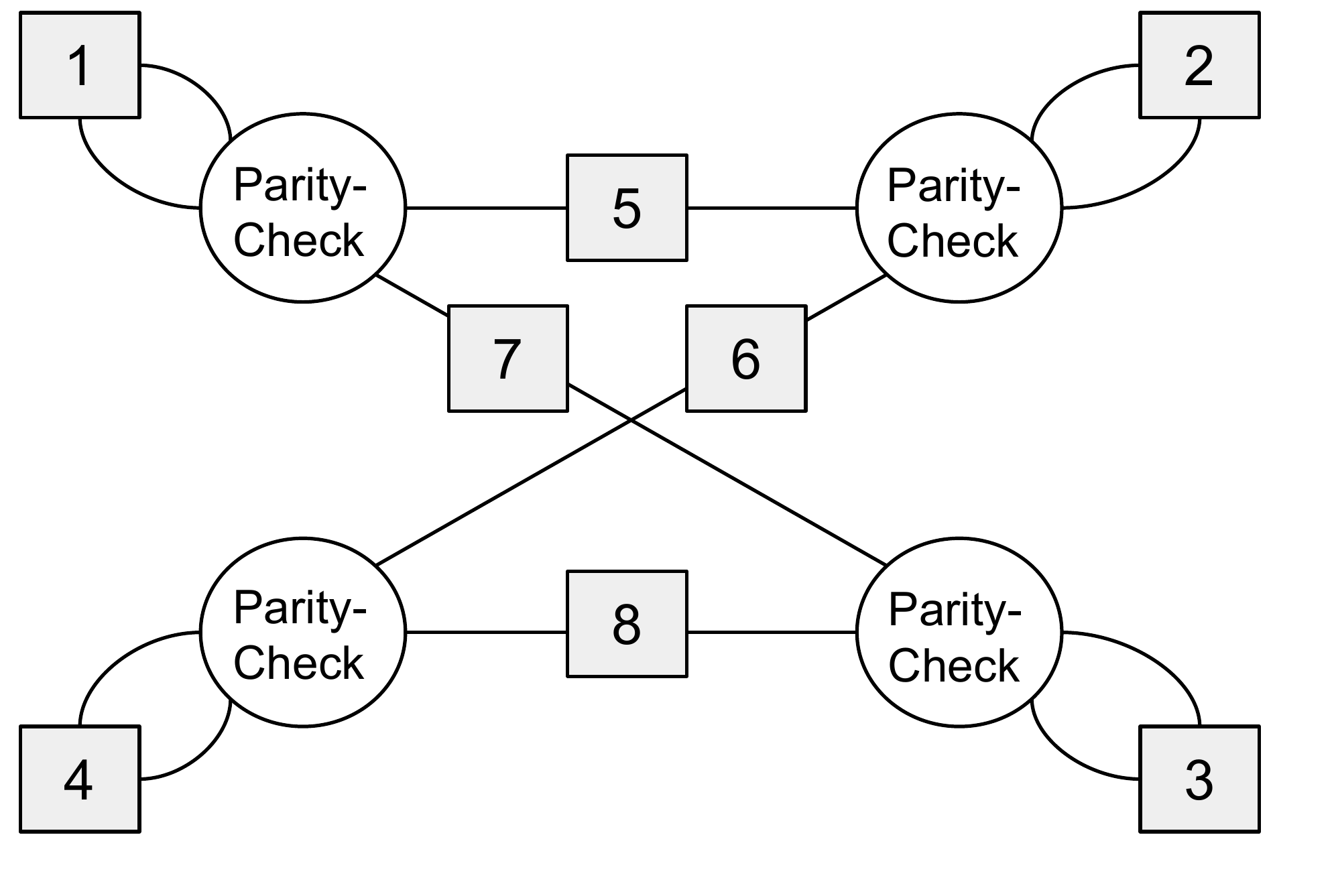}
			\caption{An example code with sequential recovery which can recover from $2$ erasures with $n=8,k=4,r=2$.}  \label{fig:turan}   
	\end{figure}      
\end{center}

The most general approach, by which we mean the approach that imposes the least constraint in terms of how recovery is to be accomplished is sequential recovery \cite{Pralalbalpvk,BalKinKum}.  An example of a code with sequential recovery is shown in Fig.~\ref{fig:turan}. In the figure, the numbers shown correspond to the indices of the $8$ code symbols.  The $4$ vertices correspond to the $4$ parity checks. Each parity check represents the equation that the sum of code symbols corresponding to the numbers attached to it, is equal to $0$. It can be seen that if the code symbols $1$ and $5$ are erased, and one chooses to decode using locality, then one must first decode code symbol $5$ before decoding symbol $1$. Hence, this code can recover sequentially from two erasures where each erasure is recovered by contacting $r=2$ code symbols with block length $n=8$ and dimension $k=4$. 

More formally a code with \textbf{sequential recovery} from $t$ erasures is an $[n,k]$ linear code over $\mathbb{F}_q$ such that an arbitrary set of $t$ erased symbols $c_{j_1},\dots,c_{j_t}$ can be recovered as follows:
\begin{eqnarray*}
	c_{j_i} = \sum_{m \in S_i} a_m c_m, \ \ a_m \in \mathbb{F}_q 
\end{eqnarray*}
where $S_i \subseteq [n]$, $|S_i| \leq r$ and $\{j_{i},j_{i+1}...,j_t\} \cap S_i = \emptyset $. The example given in the Fig.~\ref{fig:turan} corresponds to the parameter set $(n=8,k=4,r=2,t=2)$. Sequential recovery was introduced in \cite{Pralalbalpvk}, where a detailed analysis for $t=2$ case can be found. Characterization of maximum possible rate for given $n,r$ and $t=3$ can be found in \cite{SonYue}. The maximum possible rate of codes with sequential recovery for a given $r,t$ is characterized in \cite{BalKinKum}. The construction of codes having high rate but rate that is lesser in comparison to the rate of the construction in \cite{BalKinKum} can be found in \cite{SonCaiYueCaiHan,RawMazVis}.   The construction in \cite{SonCaiYueCaiHan} however has lower block length $O(r^{O(log(t))})$ in comparison to the construction in \cite{BalKinKum} that has $O(r^{O(t)})$ block length.

\subsubsection{Codes With Parallel Recovery}
If in the definition of sequential recovery, we impose the stronger requirement $\{j_{1},\dots,j_t\} \cap S_i = \emptyset $ for $1 \leq i \leq t$, in place of $\{j_{i},j_{i+1},\dots,j_t\} \cap S_i = \emptyset $ we will obtain the definition of a code with \textbf{parallel recovery}.  Thus under parallel recovery each of the $t$ erased code symbols can be recovered in any desired order. Please see \cite{PamHolOgg} for additional details on parallel recovery.  

\subsubsection{Codes With Availability} 
\begin{center}
	\begin{figure}[ht!]
		\centering    
		\includegraphics[width=2.2in]{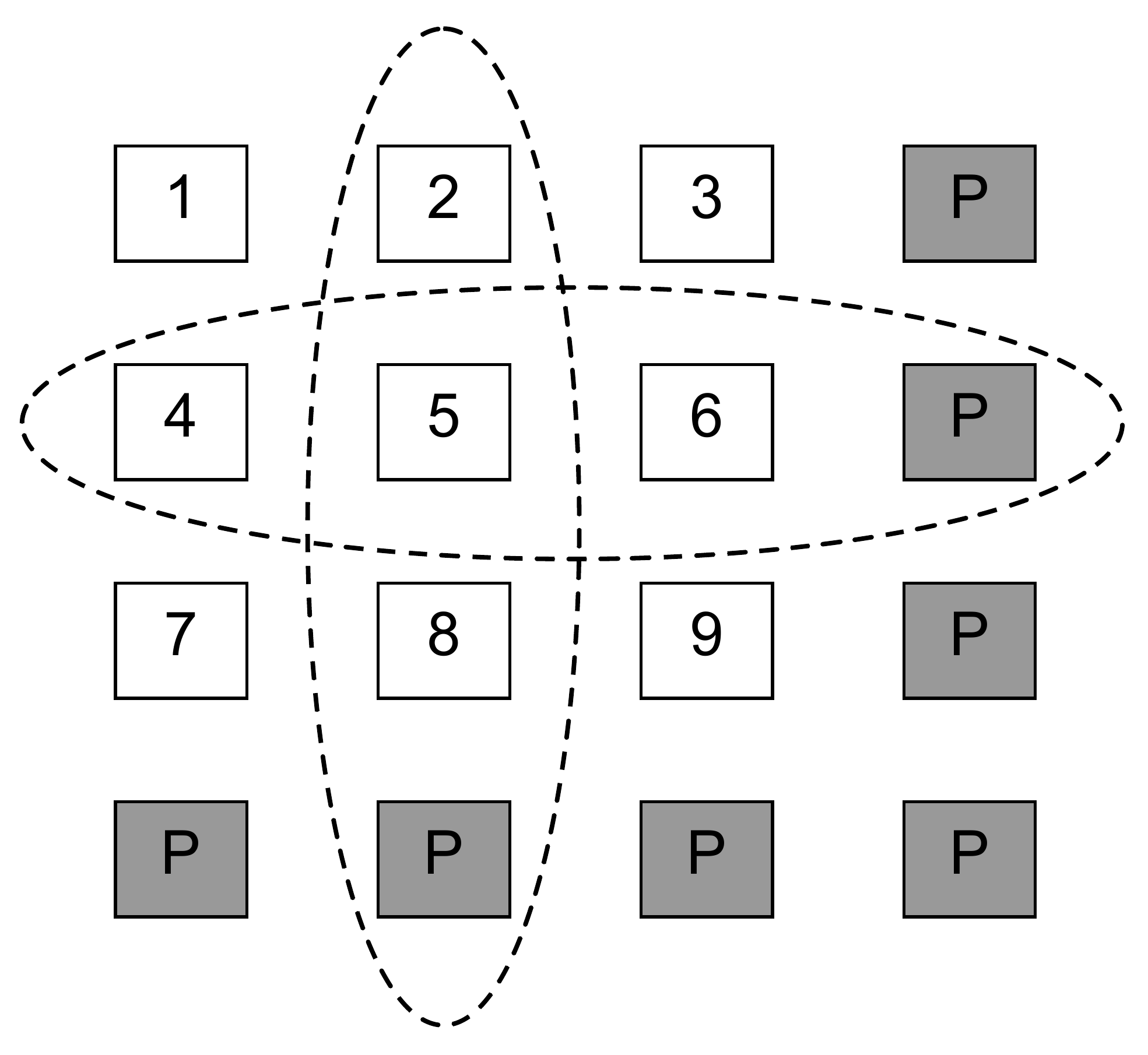}
			\caption{The binary product code as an example of a code with availability.  In this example, the code symbol 5 can be recovered either directly from the node storing it, or else by computing either the row sum or the column sum.}  \label{fig:product_code}  
	\end{figure}      
\end{center}

Codes with availability \cite{WanZha,WanZhaLiu} cater to the situation when a node containing a code symbol that it is desired to access, is unavailable as the particular node is busy serving other requests.  To handle such situations, an availability code is designed so that the same code symbol can be recovered in multiple ways, as a linear combination of a small and disjoint subset of the remaining code symbols.  The binary product code is one example of an availability code.  Consider a simple example of a product code in which code symbols are arranged in the form of an $(r+1)\times (r+1)$ array and the code symbols are such that each row and column satisfies even parity (see Fig.~\ref{fig:product_code}) for an example). Thus the code symbols within any row or column of the array sum to zero.   It follows that each code symbol can be recovered in $3$ distinct ways: directly from the node storing the code symbol or else by computing the sum of the remaining entries in either the row or the column containing the desired symbol. 

Formally, a code with \textbf{$t$-availability}, is an $[n,k]$ linear over $\mathbb{F}_q$ such that each code symbol $c_i$ can be recovered in $t$ disjoint ways as follows:
\begin{eqnarray*}
	c_{i} = \sum_{m \in S^i_j} a_m c_m, \ \ a_m \in \mathbb{F}_q 
\end{eqnarray*}
where for $1 \leq j \neq j' \leq t$, we have that 
\begin{eqnarray*}
	i \notin S^i_j, \ \ |S^i_j| \leq r \ \ \text{ and } \ \ S^i_{j} \cap S^i_{j'} = \emptyset .
\end{eqnarray*}
The sets $\{S^i_j\}$ will be referred to as recovery sets.   The example product code described above corresponds to the parameter set $(n=(r+1)^2,k=r^2,t=2)$ and the sets $S^i_1,S^i_2$ to symbols lying within the same row and column respectively.

Codes with $t$-availability can recover from $t$ erasures.  This can be seen as follows. If there are $t$ erased symbols including $c_i$, then apart from $c_i$ there are $t-1$ other erased symbols.  These however, can be present in at most $t-1$ out of the $t$ disjoint recovery sets $S^i_1,\dots,S^i_t$. Hence, there must exist at least one recovery set $S^i_j$ in which none of the erased symbols is present and this recovery set can be used to recover $c_i$. It can be verified that a code with $t$-availability is also a code that can recover for $t$ erasures in parallel. 

Upper bounds on the rate of codes with availability for a given $(r,t)$ can be found in \cite{TamoBarg_Ava, BalPVK, SwaCal}. A construction of high-rate codes with availability is presented in \cite{WanZhaLiu}.
An upper bound on the minimum distance of a code with availability that is independent of field size $q$, can be found in \cite{WanZha,TamoBarg_Ava,BalPVK}.  Field size dependent upper bounds on minimum distance can be found in \cite{HuaYakUchSie,BalPVK}. Asymptotic lower bounds for fixed $(r,t,q)$, i.e., lower bounds on rate $\frac{k}{n}$ as a function of relative minimum distance $\frac{d_{\min}}{n}$ as $n \rightarrow \infty$ for fixed $(r,t,q)$ can be found in \cite{TamoBarg_Ava,BarTamVla}.

\subsubsection{Codes With Cooperative Recovery}

In all the different types of $t$-erasure LRCs that we have encountered thus far, the constraint placed has always been on the number $r$ of unerased symbols contacted for the repair of a single erased symbol. 
In cooperative recovery, a constraint is placed instead, on the total number of unerased code symbols contacted for the recovery of all $t$ erased symbols. Formally, a code with \textbf{cooperative recovery} is an $[n,k]$ linear code over a field $\mathbb{F}_q$ such that an arbitrary set $\{c_{j_1},\dots,c_{j_t}\}$ of $t$ erased symbols  can be recovered from a set of $t$ equations as shown below:
\begin{eqnarray*}
	c_{j_i} = \sum_{m \in S} a_{m,i} c_m, \ \ a_{m, i} \in \mathbb{F}_q, i \in [t]
\end{eqnarray*}
that involve a common set $\{c_m \mid m \in S\}$ of $r$ unerased code symbols, where $S \subseteq [n]$, $|S| \leq r$. Further details including constructions and performance bounds can be found in \cite{RawMazVis}. 

\subsubsection{Codes With $(r,\delta)$ Locality} 	

The definition of an LRC required that each code symbol be part of a single parity check code of length $\leq (r+1)$. If it was required instead, that each code symbol be part of an $[r+\delta-1,r,\delta]$ MDS code, then the resultant code would be an example of a code with $(r,\delta)$ locality.   Thus each local code is stronger in terms of minimum distance, allowing local recovery from a larger number of erasures.   

More formally, a code with \textbf{$(r,\delta)$ locality} $\mathcal{C}$ over $\mathbb{F}_q$ is an $[n,k]$ linear code over $\mathbb{F}_q$ such that for each code symbol $c_i$ there is an index set $S_i \subseteq [n]$ such that $d_{\min}(\mathcal{C}|_{S_i}) \geq \delta$ and $|S_i| \leq r+\delta-1$ where $\mathcal{C}|_{S_i}$ is the restriction of the code to the coordinates corresponding to the set $S_i$.  Alternately, we may regard the code $\mathcal{C}|_{S_i}$ as being obtained from $\mathcal{C}$ by puncturing $\mathcal{C}$ in the locations corresponding to index set $[n] \setminus S_i$.  Note that an LRC code is an instance of an $(r,\delta)$ code with $\delta=2$. 

The classification of this class of codes into information symbol and all symbol $(r,\delta)$ locality codes follows in the same way as was carried out in the case of an LRC. There is an analogous minimum distance bound \cite{PraKamLalKum} given by:
\begin{eqnarray}
d_{\min}(\mathcal{C}) & \leq & (n-k+1) - \left( \left \lceil \frac{k}{r} \right \rceil -1 \right) (\delta-1). \label{eq:optimaldminrdelta}
\end{eqnarray}
A code with $(r,\delta)$ locality satisfying the above bound with equality is said to be optimal. Optimal codes with $(r,\delta)$ information symbol locality can be obtained from pyramid codes by extending the approach described in Section \ref{pyramid} and splitting a larger number $(\delta-1)$ of parity columns in the generator matrix of a systematic RS code. Optimal codes with $(r,\delta)$ ASL can be obtained by employing the construction in \cite{TamBar_LRC} as described in Section \ref{TamoBarg} for the case when $(r+\delta-1) \vert n$ and $q=O(n)$. Optimal $(r,\delta)$ cyclic codes with $q=O(n)$ can be found in \cite{CheXiaHaoFu} for the case when $(r+\delta-1) \vert n$.  A detailed analysis as to when the upper bound on minimum distance appearing in \eqref{eq:optimaldminrdelta} is achievable can be found in  \cite{SonDauYueLi}. Characterization of binary codes achieving the bound in \eqref{eq:optimaldminrdelta} with equality can be found in \cite{HaoXiaChe}.  A field size dependent upper bound on dimension $k$ for fixed $(r,\delta,n,d_{\min})$ appears in \cite{AgaBarHuMazTam}. Asymptotic lower bounds for a fixed $(r,\delta,q)$ i.e., lower bounds on rate $\frac{k}{n}$ as a function of relative minimum distance $\frac{d_{\min}}{n}$ as $n \rightarrow \infty$ for a fixed $(r,\delta,q)$ can be found in  \cite{BarTamVla}.

\subsubsection{Hierarchical Codes} 	

\begin{figure}[h!]
	\begin{center}
		\subfigure[LRC]{
			\includegraphics[width=2.8in]{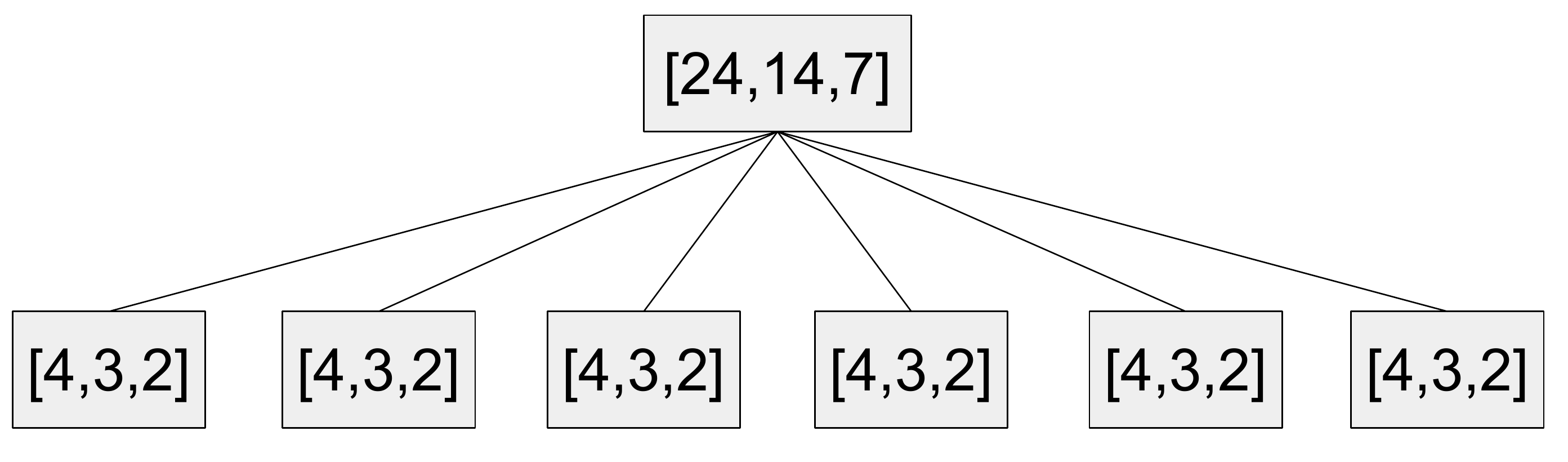} }
		\subfigure[Hierarchical LRC]{
			\includegraphics[width=2.8in]{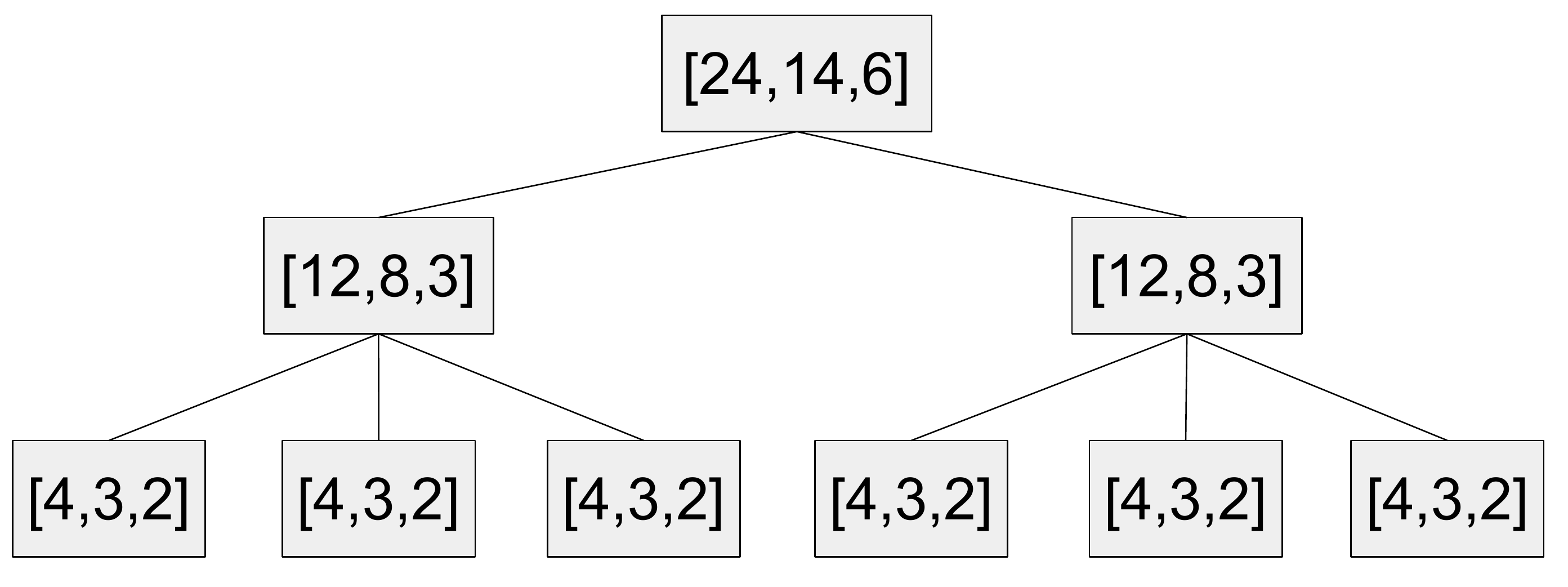} }
				\caption{The code on the left is an LRC in which each code symbol is protected by a $[n=4,k=3,d_{\min}=2]$ local code and each local code is contained in an overall $[24,14,7]$ global code. In the hierarchical locality code appearing to the right, each local code is a part of a $[12,8,3]$ so-called middle code, and the middle codes in turn, are contained in an overall $[24,14,6]$ global code.} \label{fig:hierarchical} 
	\end{center}
\end{figure}

From a certain perspective, the idea of an LRC is not scalable. Consider for instance, a $[24,14]$ linear code which is made up of the union of $6$ disjoint $[4,3]$ local codes (see the left side of Fig.~\ref{fig:hierarchical}).  These local codes are single parity check codes and ensure that the code has locality $3$.  However if there are $2$ or more erasures within a single local code, then local recovery is no longer possible and one has to resort to decoding the entire code as a whole to recover the two erasures.  Clearly, this problem becomes more acute as the block length $n$ increases. One option to deal with this situation, would be to build codes with $(r,\delta)$ locality but even in this case, if there are more than $(\delta-1)$ erasures within a local code, local decoding is no longer possible.  Codes with hierarchical locality \cite{SasAgaKum_loc,BalBarVal} (see Fig.~\ref{fig:hierarchical} (right)) seek to overcome this by building a hierarchy of local codes having increasing block length, to ensure that in the event that a local code at the lowest level is overcome by a larger number of erasures than it can handle, then the local code at the next level in the hierarchy can take over.  As one goes up the hierarchy, both block length and minimum distance increase.  An example hierarchical code is presented in Fig.~\ref{fig:hierarchical}. 

\subsection{Maximally Recoverable Codes}

Let $\mathcal{C}$ be an $[n,k]$ linear code with $(r,\delta)$ locality such that every local code has disjoint support and where further, each local code is an $[r+\delta-1,r]$ MDS code.
Let $E \subseteq [n]$ be formed by picking $(\delta - 1)$ coordinates from each of the local codes within $\mathcal{C}$. Then $\mathcal{C}$ is said to be \textbf{maximally recoverable} (MR) \cite{GopHuaJenYek} if the code obtained by puncturing $\mathcal{C}$ on coordinates defined by $E$ is an MDS code. An MR code can correct all possible erasure patterns that are information-theoretically correctable given the locality constraints. MR codes were originally introduced as partial MDS codes in \cite{Bla_Haf_Het}. The notion of maximal recoverability finds particular application in the design of sector-disk codes \cite{PlaBla} that are used in RAID storage systems to combat simultaneous sector and disk erasures.

\section{Locally Regenerating Codes}
\begin{center}
	\begin{figure}[h!]
		\centering
		\includegraphics[width=5.5in]{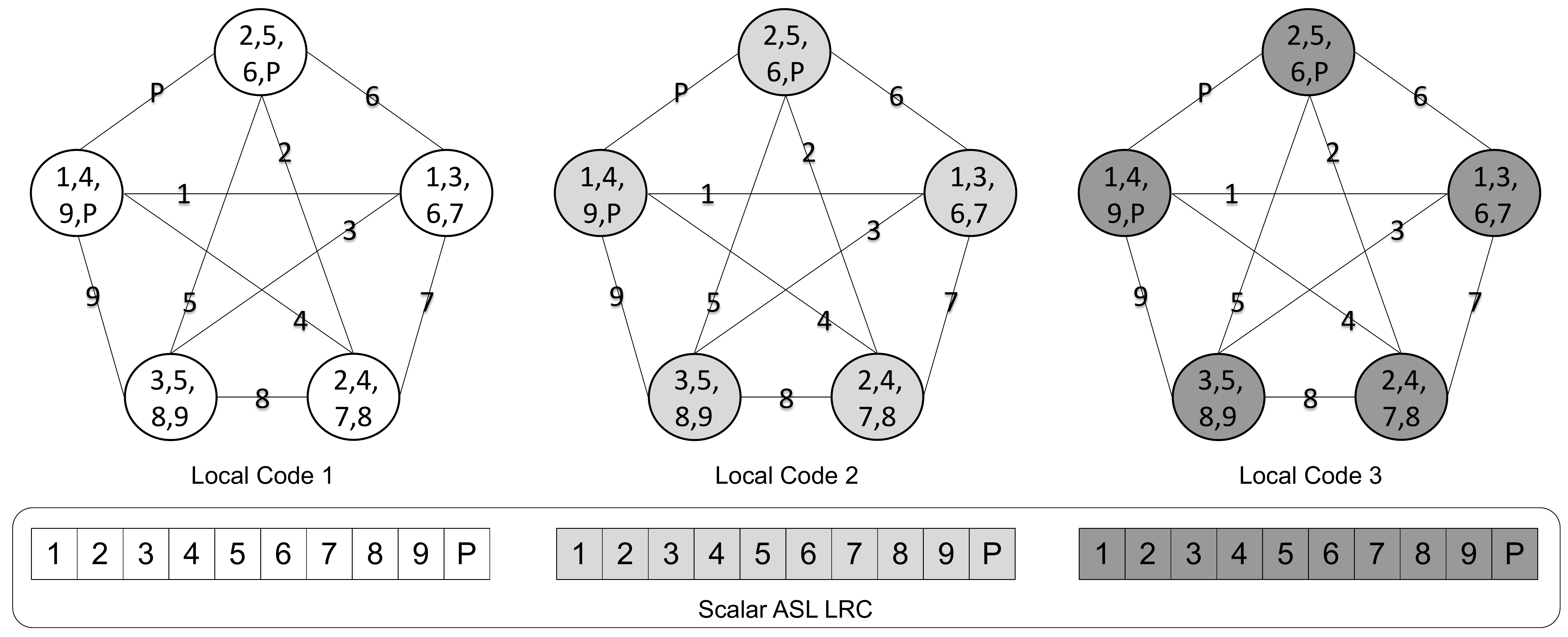}  
			\caption{An LRGC in which each of the three local codes is a pentagon MBR code. The set of $30$ scalar symbols that make up the LRGC form a scalar, ASL LRC in which there are three disjoint local codes, each of block length $(r+1)=10$.   The contents of each of the three pentagons are obtained from the $10$ scalar symbols making up the respective local code by following the same procedure employed to construct a pentagon MBR code from a set of $10$ scalar symbols that satisfy an overall parity check.}
		\label{fig:LRGC}    
	\end{figure}      
\end{center}

We have seen earlier that while RGCs minimize repair bandwidth, LRCs minimize the repair degree. Locally Regenerating Codes (LRGCs) \cite{RawatKoyluSilberVish,KamPrakLalKumLRGC14} are codes which simultaneously possess low repair bandwidth as well as low repair degree. LRGCs are perhaps best viewed as vector codes with locality, in which the local codes are themselves RGCs. In Fig. \ref{fig:LRGC}, we illustrate an LRGC where each local code is a repair-by-transfer, pentagon MBR code.

\section{Efficient Repair of Reed-Solomon Codes} \label{Sec:RS_repair}

In an $[n,k]$ MDS code, each code symbol is traditionally considered as an indivisible unit over $\mathbb{F}_q$. As any $k$ code symbols of an MDS code form an information set, the conventional approach to the recovery of an erased code symbol, is to access an arbitrary set of $k$ other code symbols, use these to decode the code and in this way, recover the erased symbol.   In the context of a distributed storage system, where each code symbol of a codeword is stored in a distinct node, this implies a repair bandwidth which is $k$ times the amount of data stored in a failed node.  The first step in developing a more efficient repair strategy in the case of an MDS code, is to employ a finite field $\mathbb{F}_q$ that is a degree-$t$ extension of a base field $\mathbb{B}$. Thus, if $|\mathbb{B}|=p$, where $p$ is some prime power, we will have $|\mathbb{F}_q|=q=p^t$.  In this setting, the authors of \cite{ShanPapDimCai}, then took the important next step of regarding each code symbol as a vector over the base field $\mathbb{B}$\ and showing that the repair bandwidth can be reduced by carrying out repair operations over the base field.  The paper \cite{ShanPapDimCai} dealt with the specific case $n-k=2$, where only the repair of systematic nodes was considered. This approach was subsequently generalized in \cite{GuruWoot} to present an efficient all-node-repair scheme for Generalized Reed-Solomon (GRS) (see \cite[Ch.~10]{mcwslo}) codes as described below. \\
\textbf{GRS Codes}: Let $\Theta:=\{\theta_i\}_{i=1}^n\subseteq\mathbb{F}_q$ denote a subset of $\mathbb{F}_q$\ of size $n$.   Let $\mathcal{F}$, ${\mathcal H}$\ denote the set of all polynomials in $\mathbb{F}_q[x]$ having degree bounded above by $(k-1)$ and $(n-k-1)$ respectively.  Each codeword in an $[n,k]$ GRS code $\mathcal{C}_\text{GRS}$ is obtained by evaluating a polynomial in ${\mathcal F}$, along with scaling coefficients $\{u_i\}_{i=1}^n\subseteq \mathbb{F}_q^{*}$, at the elements of $\Theta$, i.e., 
\begin{equation}\nonumber
\mathcal{C}_\text{GRS}=\{(u_1f(\theta_1),u_2f(\theta_2),\ldots,u_nf(\theta_n)) \mid f \in {\mathcal F}\}.
\end{equation}
The dual code of ${\mathcal F}$ is then of the form:
\begin{equation}\nonumber
\mathcal{C}_\text{GRS}^\perp=\{(v_1h(\theta_1),v_2h(\theta_2),\ldots,v_nh(\theta_n)) \mid h \in {\mathcal H}\}.
\end{equation}
Like the $\{u_i\}$, the $\{v_i\}_{i=1}^n\subseteq\mathbb{F}_q^*$ are also a set of scaling coefficients. The scaling coefficients $\{u_i\},\{v_j\}$ do not however, play any role in determining the repair bandwidth and for this reason, in the text below, we assume all the scaling coefficients $u_i,v_j$ to equal $1$.  

\textbf{Trace Function and Trace-Dual Basis}:   The trace function $T:\mathbb{F}_q\rightarrow\mathbb{B}$ is given by:
\begin{equation}\nonumber
T(x)=\sum_{i=0}^{t-1}x^{p^i}, 
\end{equation}
where $x\in\mathbb{F}_q$. For every basis $\Gamma=\{\gamma_1,\gamma_2,\ldots,\gamma_t\}$, of $\mathbb{F}_q$ over $\mathbb{B}$, there exists a second basis $\Delta=\{\delta_1,\delta_2,\ldots,\delta_t\}$ termed the trace-dual basis satisfying:
\begin{equation}\nonumber
T(\gamma_i\delta_j) = \left\{\begin{array}{lr}
1, & i=j\\
0, & \text{else.}
\end{array}\right. 
\end{equation}
It can be verified that each element $x \in \mathbb{F}_q$ has the basis expansion:
\begin{equation}\nonumber
x=\sum_{i=1}^t T(x\gamma_i)\delta_i.
\end{equation}
Thus given $\{T(x\gamma_i)\}_{i=1}^t$, the element $x$ can be uniquely recovered.

\textbf{Node Repair via the Dual Code}: Recall that $\mathcal{C}_\text{GRS}$ and its dual $\mathcal{C}_\text{GRS}^\perp$ are scaled evaluations of polynomials of degree at most $k-1$ and at most $n-k-1$, respectively. Hence for $f,h \in {\mathcal F}, {\mathcal H}$ respectively, we have $\sum_{i=1}^n f(\theta_i)h(\theta_i)=0$ (pretending that each $u_i$ and each $v_j$ equals $1$ for reasons explained earlier).  Let us assume that code symbol $f(\theta_i)$ has been erased. We have:
\begin{equation}\nonumber
f(\theta_i)h(\theta_i)=-\sum_{j=1,j\neq i}^n f(\theta_j)h(\theta_j).
\end{equation}
Thus,
\begin{equation}\label{eq:GRS_repair_equation}
T(f(\theta_i)h(\theta_i))=-\sum_{j=1,j\neq i}^n T(f(\theta_j)h(\theta_j)).
\end{equation}

Next, let us assume that it is possible to select a subset $\mathcal{H}_i$ of $\mathcal{H}$ 
in such a way that $\{h(\theta_i)\}_{h\in\mathcal{H}_i}$ forms a basis for $\mathbb{F}_q$\ over $\mathbb{B}$. It follows from \eqref{eq:GRS_repair_equation} and the existence of a trace-dual basis that $f(\theta_i)$ can be recovered from the set $\left\{\sum_{j=1,j\neq i}^n T(f(\theta_j)h(\theta_j))\right\}_{h\in\mathcal{H}_i}$. In \cite{GuruWoot}, the authors carefully choose the subsets $\{\mathcal{H}_i\}_{i=1}^n$ so as to not only satisfy the above basis requirement, but also reduce the repair bandwidth associated with the recovery of $f(\theta_i)$ via \eqref{eq:GRS_repair_equation}.

{\bf The Repair Scheme in \cite{GuruWoot}}: Let $n-k\geq p^{t-1}$ for a GRS code ${\mathcal C}$. Let $\Gamma=\{\gamma_1,\gamma_2,\ldots,\gamma_t\}$ be a basis for $\mathbb{F}_q$\ over $\mathbb{B}$.  Each codeword in ${\mathcal C}$ corresponds to the scaled evaluation at the elements in $\Theta$, of a polynomial $f\in {\mathcal F}$. With respect to the scheme for failed-node recovery described above, consider the set 
\begin{equation*}\label{eq:parity_poly_construction}
\mathcal{H}_i=\left\{\frac{T\big(\gamma_j(x-\theta_i)\big)}{(x-\theta_i)}\right\}_{j=1}^t.
\end{equation*}
It is straightforward to verify that $\{h(\theta_i)\}_{h\in\mathcal{H}_i}\equiv\Gamma$ 
and for $j\neq i$, $\{h(\theta_j)\}_{h\in\mathcal{H}_i}$ is a set consisting of scalar multiples (over $\mathbb{B}$) of $\frac{1}{\theta_j-\theta_i}$. Hence, from the $\mathbb{B}$-linearity of the trace function $T$, it is possible to compute all elements in the set $\{T(f(\theta_j)h(\theta_j))\}_{h\in\mathcal{H}_i}$ from $T(\frac{f(\theta_j)}{\theta_j-\theta_i})$.   Clearly, in order for the replacement node to be able to compute $\left\{\sum_{j=1,j\neq i}^n T(f(\theta_j)h(\theta_j))\right\}_{h\in\mathcal{H}_i}$, each node-$j$ ($j\neq i$) needs only provide the single symbol $T(\frac{f(\theta_j)}{\theta_j-\theta_i})\in\mathbb{B}$.   This results in a repair bandwidth of $n-1$ symbols over $\mathbb{B}$ to recover each $f(\theta_i)$. In contrast, as noted earlier, the traditional approach for recovering a code symbol incurs a repair bandwidth of $k$ symbols over $\mathbb{F}_q$ or equivalently, $kt$ symbols over $\mathbb{B}$. 

There has been much subsequent work on the repair of codes, dealing with issues such as repairing RS codes in the presence of multiple erasures, achieving the cut-set bound on node repair, enabling optimal access etc.  The reader is referred to \cite{DauDuuKiaMil,TamoYeBarg19,ChenYeBarg} and the references therein for details.  RS repair schemes specific to the $[n=14, k=10]_{q = 256}$ RS code employed by HDFS have been provided in \cite{DuurDau}.

\section{Codes for Distributed Storage in Practice}
Given the clear-cut, storage-overhead advantage that erasure codes provide over replication, popular distributed systems such as Hadoop, Google File System (GFS), Windows Azure, Ceph and Openstack have enabled support for erasure codes within their systems. These erasure coding options were initially limited to RS codes.  It was subsequently realized that the frequent node-repair operations taking place in the background and the consequent network traffic, and helper-node distraction, were hampering front-end operations. This motivated the development of the RGCs, LRCs and the improved repair of RS codes.  As noted in Section~\ref{sec:Azure}, LRCs are very much a part of the Windows Azure cloud-storage system.  
Hadoop EC has made Piggybacked RS codes available as an option.  Both LRC and MSR (Clay) codes are available as erasure coding options in Ceph. 

\bibliographystyle{IEEEtran}
\bibliography{storage_bib}

\end{document}